\documentclass[floatfix,prd,twocolumn,letterpaper,lengthcheck,superscriptaddress,showpacs,amssymb,amsmath,amsfonts,aps,altaffilletter,nofootinbib,nopreprintnumbers,longbibliography]{revtex4-2}

\usepackage[dvipsnames]{xcolor}
\usepackage{hyperref}
\usepackage{amsmath}
\usepackage{amsthm}
\usepackage{multirow}
\usepackage{graphicx}
\usepackage{caption}
\usepackage{subcaption}
\usepackage{placeins}
\usepackage{orcidlink}

\usepackage{mathrsfs}
\DeclareSymbolFontAlphabet{\mathrsfs}{rsfs}

\newcommand{\scrM}{\mathrsfs{M}}

\def\MM{\mathcal{M}} 

\newcommand{\Surf}{\ensuremath{\mathcal{S}}}
\def\insubscript{\rm inner}
\def\outsubscript{\rm outer}
\newcommand{\Sout}{\Surf_{\outsubscript}}   
\newcommand{\Sin}{\Surf_{\insubscript}}     
\newcommand{\Sone}{\Surf_{1}}               
\newcommand{\Stwo}{\Surf_{2}}               
\newcommand{\Sonetwo}{\Surf_{1,2}}

\newcommand{\HH}{\ensuremath{\mathcal{H}}}

\def\tname{t} 
\def\ttouch{\tname_{\rm touch}}
\def\tbifurcate{\tname_{\rm bifurcate}}

\newcommand{\stilde}{\tilde{s}}

\newcommand{\RR}{\ensuremath{\mathcal{R}}}           
\newcommand{\Rin}{\RR_{\insubscript}}                
\newcommand{\Rone}{\RR_1}                
\newcommand{\Rtwo}{\RR_2}                

\newcommand{\Rinone}{\RR^{(1)}_{\insubscript}}       
\newcommand{\Rintwo}{\RR^{(2)}_{\insubscript}}       
\newcommand{\thneck}{\ensuremath{\theta_{\star}}} 
\newcommand{\zetaneck}{\ensuremath{\zeta_{\star}}} 
\newcommand{\sneck}{\ensuremath{\stilde_{\star}}} 

\begin{document}

\title[]{Cusp Formation in Merging Black Hole Horizons}
\author{Shilpa Kastha}
\affiliation{
  Saha Institute of Nuclear Physics, 1/AF Bidhannagar, Kolkata 700064, India
}
\affiliation{
  Homi Bhabha National Institute, Training School Complex, Anushaktinagar, Mumbai 400094, India
}
\affiliation{
  Max-Planck-Institut f\"ur Gravitationsphysik (Albert Einstein Institute),
  Callinstr. 38, 30167 Hannover, Germany
}
\affiliation{
  Leibniz Universit\"at Hannover, 30167 Hannover, Germany
}
\affiliation{
  Niels Bohr International Academy, Niels Bohr Institute,
  Blegdamsvej 17, 2100 Copenhagen, Denmark
}

\author{Stamatis Vretinaris
  \orcidlink{0000-0001-7575-813X}
}
\affiliation{
  Max-Planck-Institut f\"ur Gravitationsphysik (Albert Einstein Institute),
  Callinstr. 38, 30167 Hannover, Germany
}
\affiliation{
  Leibniz Universit\"at Hannover, 30167 Hannover, Germany
}
\affiliation{
  Institute for Mathematics, Astrophysics and Particle Physics, Radboud University,
  Heyendaalseweg 135, 6525 AJ Nijmegen, The Netherlands
}

\author{Daniel Pook-Kolb}
\affiliation{
  Max-Planck-Institut f\"ur Gravitationsphysik (Albert Einstein Institute),
  Callinstr. 38, 30167 Hannover, Germany
}
\affiliation{
  Leibniz Universit\"at Hannover, 30167 Hannover, Germany
}
\affiliation{
  Institute for Mathematics, Astrophysics and Particle Physics, Radboud University,
  Heyendaalseweg 135, 6525 AJ Nijmegen, The Netherlands
}

\author{Badri Krishnan}
\affiliation{
  Max-Planck-Institut f\"ur Gravitationsphysik (Albert Einstein Institute),
  Callinstr. 38, 30167 Hannover, Germany
}
\affiliation{
  Leibniz Universit\"at Hannover, 30167 Hannover, Germany
}
\affiliation{
  Institute for Mathematics, Astrophysics and Particle Physics, Radboud University,
  Heyendaalseweg 135, 6525 AJ Nijmegen, The Netherlands
}


\begin{abstract}

  An important question in binary black hole mergers is to connect
  properties of the remnant black hole to those of the two initial
  black holes.  These properties include not only the final mass and
  spin of the remnant, but also higher multipoles and answers to other
  questions such as, for a given initial configuration, which
  quasi-normal modes of the final black hole are excited, and what are
  the amplitudes of these modes?  Such questions have thus far been
  primarily addressed through a study of the emitted gravitational
  wave signal.  In this paper we consider a different alternative,
  namely using quasi-local black hole horizons themselves to establish
  the link between the initial and final states.  Recent work has
  elucidated the behavior of black hole horizons in a merger.  Cusps
  forming in such otherwise smoothly evolving horizons have been shown
  to play a central role in connecting the two initially separate
  black holes with the final remnant.  In the present work, we will
  discuss from a numerical perspective how such cusps form in detail
  for the head-on collision of two non-spinning black holes.  We show
  how the mass and higher mass multipole moments behave at the cusp
  and suggest a phenomenological model.

\end{abstract}

\maketitle

\section{Introduction}
\label{sec:intro}

For almost its entire lifetime, any black hole in our universe is
described with great accuracy by a particular solution to the Einstein
equations, namely the Kerr solution. Thus, given that the Kerr
solutions are a 2-parameter family of solutions, black holes and the
spacetime geometry in their vicinity are determined by just two
parameters. These are, of course, the mass $M$ and spin angular
momentum $\mathbf{S}$ (often expressed in terms of the dimensionless
spin parameter $\chi := |\mathbf{S}|/M^2$ which satisfies $\chi <
  1$). All higher multipole moments are determined by these two
parameters.  Even for black holes with external perturbations such
accretion disks, or a distant binary companion, the black holes are
still well approximated as Kerr black holes.  If necessary, the higher
order effects of such external influences can be accurately modeled by
linear perturbations of the Kerr solution.  All of these features make
black holes rather distinct from other stars where the higher
multipole moments are not determined uniquely by the mass and angular
momentum.

There is however one clear case where the Kerr solution does not
suffice, namely when two black holes get sufficiently close to each
other and coalesce to yield a single remnant black hole.\footnote{Even
  for black hole mergers, black hole perturbation theory retains a lot
  of its utility when the binary system is highly asymmetric
  \cite{Emparan:2017vyp}.} For the merger of two roughly comparable
mass black holes, non-perturbative effects must necessarily be taken
into account.  Even though the merger lasts for a very short duration
compared to the lifetime of a typical black hole, the black hole mass
and spin generally change appreciably during the merger. One may
therefore expect the distribution of black hole masses and spins in
our universe to be significantly impacted by binary mergers. It is
currently not possible to obtain a full analytic solution of the
Einstein equations for a binary black hole merger, but the problem can
be addressed numerically.  Over the past two decades, numerical
relativity has matured and it is now possible to obtain accurate
numerical solutions to the binary black hole problem (see,
e.g., \cite{Sperhake:2014wpa,Centrella:2010mx}).

One can view a binary black hole merger as a process whereby two Kerr
black holes with masses $(M_1,M_2)$, spin angular momenta
$(\mathbf{S}_1,\mathbf{S}_2)$, and linear momenta
$(\mathbf{P}_1,\mathbf{P}_2)$ merge to yield a remnant Kerr black hole
with mass $M_f$, spin $\mathbf{S}_f$ and momentum $\mathbf{P}_f$.  If
the two initial black holes are gravitationally bound in a binary
system, then $\mathbf{P}_1$ and $\mathbf{P}_2$ are not independent and
would be determined by properties of the binary system such as the
initial separation, eccentricity etc.  We can encapsulate this process
as a mapping from an initial to a final configuration:
\begin{equation}\label{eq:map}
  \left\{ \begin{matrix} (M_1,\mathbf{S}_1,\mathbf{P}_1) \\ (M_2,\mathbf{S}_2,\mathbf{P}_2) \end{matrix}\right\}
  \rightarrow (M_f,\mathbf{S}_f,\mathbf{P}_f)\,.
\end{equation}
Determining the remnant black hole parameters following such a mapping
is useful for several reasons. In the case of supermassive black
holes, the final parameters would be linked to several astrophysical
observables such as the location of the central black hole within the
host galaxy, the orientation of jets in the case of active galactic
nuclei, the velocity dispersion of stars within the central galactic
bulge etc; see, e.g., \cite{Merritt2005LRR}.  For stellar mass black
holes such as those observed by the LIGO, Virgo and KAGRA
observatories, the distribution of black hole parameters and our
inferences regarding the various formation channels of binary black
hole systems would depend on the above mapping; see,
e.g., \cite{PhysRevD.100.043027}. Finally, the spectrum of quasi-normal
modes emitted by the remnant black holes, and the associated tests of
general relativity
\cite{Dreyer:2003bv,Berti:2016lat} is also
determined by the mass and spin of the remnant black hole.  Moreover,
a detailed understanding of the \emph{approach} to the final state
will enable us to determine precisely which quasi-normal modes will be
excited for a given initial configuration.

Several approaches have been previously employed in the literature to
determine the above map from the initial to final states, all relying
on numerical relativity results to varying degrees.  The first is a
phenomenological approach based on a large number of numerical
relativity simulations of binary black hole mergers with a wide
variety of initial conditions
\cite{Zlochower:2015wga,Healy:2014yta,Lousto:2013wta,Hofmann:2016yih,Tichy:2008du,Barausse:2009uz,Rezzolla:2008sd,Rezzolla:2007rd,Varma:2018aht}.
A systematic approach which exploits the underlying symmetries in the
problem and employs a power series expansion in the spins is given in
\cite{Boyle:2007ru,Boyle:2007sz,Cao:2011zza}. A third approach relies
on employing suitably accurate models for the emitted gravitational
wave signal.  Given a gravitational waveform, one can calculate the
radiated fluxes of energy and angular momentum which, given the total
energy and angular momentum at any given time, can be used to compute
the remnant mass and spin.  This approach has been used thus far
within the Effective-One-Body formalism
\cite{Damour:2007cb,Damour:2013tla}.

In this paper we initiate an entirely different approach to this
problem, namely by tracking the black hole horizon dynamics and
multipole moments all the way through the merger.  As is conventional
and practically useful in numerical relativity, we shall use the
notion of quasi-local horizons (QLHs), which are 3-surfaces obtained
by the time evolution of ``apparent horizons'' or more correctly
marginally outer trapped surfaces (MOTSs)
\cite{Ashtekar:2004cn,Krishnan:2007va,Ashtekar:2003hk,Ashtekar:2025wnu,Booth:2005qc}.
These notions will be defined more precisely below; for now, we note
that QLHs differ significantly from event horizons only in the
dynamical merger regime where they lie behind the event horizon.  We
apply this framework specifically to the head-on collision of two
non-spinning black holes (though we expect the method to be
generalizable for general configurations).  This will give us the
masses, and higher multipole moments of the two individual black holes
as functions of time up to the merger, followed by a discontinuous
jump at the merger, and then finally the subsequent evolution of the
parameters of the remnant black hole as it settles down to its final
state.  The dominant effect, as we shall see, is the discontinuous
jump at the merger.  It turns out (as verified in numerical
simulations), that masses and spin \emph{magnitudes} do not change
appreciably during the inspiral regime. If the initial spins happen to
be mis-aligned with the orbital angular momentum, then the system can
exhibit precession which modifies the direction of the two spins, but
the spin magnitudes themselves do not vary appreciably (see,
e.g., \cite{Apostolatos:1994mx}).  Similarly, the late ringdown is well
modeled as a perturbed Kerr black hole, whence the mass and spin of
the remnant do not change appreciably in this regime as well.  It is
in fact in the comparatively short merger regime where the largest
variations in mass and spin occur.  Consistent with this observation,
it is here that the black hole horizons absorb the most infalling
radiation leading to significant area increase and the consequent
variations in the physical parameters.

Horizon measures for black hole mass and angular momentum are used
routinely in numerical simulations \cite{Dreyer:2002mx}.  However,
this has so far been typically used either for the two initial black
holes (often in the initial data) or for the remnant black hole at
late times.\footnote{These horizon measures are in fact used in most
  of the works referenced above for constructing the map of
  Eq.~(\ref{eq:map}).}  \footnote{While there are some subtleties in
  comparing the remnant parameters obtained from the horizon measures
  versus those obtained from the waveform and other asymptotic data
  \cite{Iozzo:2021vnq}, the mappings obtained from the two methods
  should, in principle, be equivalent.}  The challenge here is to do
this at the merger.

As we shall describe in detail below, this involves highly distorted
horizons and the formation of cusps which are challenging to locate
numerically.  Over the past decade, accurate numerical methods have
been developed and applied in binary black hole simulations for
addressing this challenge
\cite{Pook-Kolb:2018igu,PhysRevD.100.084044,pook_kolb_daniel_2019_2591105}.
The case of head-on collisions has been studied in great detail and
the process by which two horizons merge and eventually form a single
horizon is now well understood
\cite{PhysRevLett.123.171102,pook-kolb2020I,pook-kolb2020II,Booth:2021sow,Pook-Kolb:2021jpd,Pook-Kolb:2021gsh}.
Though we shall not pursue this question in this paper, among the
physical applications of this work will be a quantitative
understanding of how the remnant black hole approaches equilibrium and
its relation to quasi-normal modes
\cite{Gupta:2018znn,Mourier:2020mwa,Khera:2023oyf}.  Can we predict,
based on the initial configuration, specifically which quasi-normal
modes are excited, and what are the amplitudes of these modes?  As
indicated earlier, such a prediction would be useful for the program
of black hole spectroscopy and for other tests of general relativity
\cite{Berti:2016lat}.  There are several works which address this
question through a detailed study of the emitted gravitational wave
signal. The present work suggests an alternative based upon the
horizons themselves.  How is it possible to relate the gravitational
wave signal to the horizons which are, after all, inside the event
horizon and thus causally disconnected from the wave zone where
gravitational wave observations are made?  The reasoning rests on the
following three considerations:
\begin{enumerate}
  \item[a)] First, there is now growing evidence that the gravitational wave
    signal observed in the wave zone is closely connected to the
    \emph{infalling} radiation at the horizon. This was proposed in
    \cite{Rezzolla:2010df,Jaramillo:2012rr,Jaramillo:2011rf}.  The
    heuristic explanation is that both the infalling flux at the horizon
    and outgoing waves that we observe have the same source, namely by
    the complicated non-linear dynamical spacetime outside the
    horizons. Thus, while the horizon dynamics is itself not the source
    of observed signals, the two are strongly correlated.  Since this
    initial suggestion, there are several numerical studies providing
    detailed evidence for the existence of such correlations
    \cite{Gupta:2018znn,Khera:2023oyf,Mourier:2020mwa,Khera:2023oyf,Chen_2022}.

  \item[b)] Second, it is understood how a given quasi-local horizon
    (QLH), i.e. a 3-surface foliated by Marginally Outer Trapped
    Surfaces (MOTSs), evolves in time. This includes fluxes across the
    QLH which leads to changes in mass, angular momentum and higher
    multipoles.  This includes ``physical process'' versions of the laws
    of black hole mechanics
    \cite{Ashtekar:2003hk,Ashtekar:2004cn,Gourgoulhon:2005ch,Ashtekar:2021kqj},
    and the development of geometric notions of time evolution on a QLH
    \cite{Ashtekar:2013qta,Chen_2022}.\footnote{There remain still some
      gaps in this formulation.  Examples are the technical details of
      how this extends when we transition between a spacelike QLH to
      one with a mixed signature.}

  \item[c)] Point (b) above only applies to the smooth part of the
    evolution; it does \emph{not} apply to the singular behavior at the
    merger. There is so far no study, either analytical or numerical, in
    this direction. The goal of this paper is to address this missing link
    in the case of a head-on merger of two non-spinning black holes.
\end{enumerate}
It should now be clear that if all three points enumerated above are
achieved, then we can establish a link between the progenitor and
remnant black holes.  We would thus have an alternative, and
complementary formalism to the gravitational wave signal itself.  In
this paper, we will consider the same simulation as in
Refs.~\cite{PhysRevLett.123.171102,PhysRevD.100.084044}, i.e. head-on
collisions of two non-spinning black holes.  We expect similar
considerations should hold also for more generic initial data with say
spinning black holes.

The rest of this paper is structured as follows. Basic notions and
definitions, and our numerical set-up are introduced in
Sec.~\ref{sec:basics}.  Sec.~\ref{sec:ricci} calculates geometrical
quantities numerically on various horizons and elucidates its
properties.  Sec.~\ref{sec:multipoles} studies the jump in the mass
and higher mass multipoles across the merger.  Finally
Sec.~\ref{sec:conclusions} presents a summary and concluding remarks.

\section{Basic Notions: Quasi-local horizons, their mergers, and
  multipole moments}
\label{sec:basics}

\subsection{Quasi-local horizons}
\label{subsec:qlh}

Let $(\MM, g_{ab})$ be a $4$-dimensional spacetime foliated by
spacelike Cauchy surfaces $(\Sigma, h_{ij}, K_{ij})$ with Riemannian
$3$-metric $h_{ij}$ and extrinsic curvature $K_{ij}$.  A smooth closed
spacelike $2$-surface $\Surf\subset\Sigma$ with a Riemannian 2-metric
$q_{ab}$ is equipped with two future pointing null normal directions.
We assume it is possible to assign an outward direction and let
$\ell^a$ and $n^a$ be future pointing outgoing and ingoing null
normals, respectively. These can be rescaled with positive definite
functions and we cross normalize these via $\ell \cdot n = -1$.
Congruences of null geodesics starting in the $\ell^a$ or $n^a$
directions then have expansions
\begin{equation}\label{eq:expansion}
  \Theta_{(\ell)} = q^{ab} \nabla_a \ell_b \,,
  \qquad
  \Theta_{(n)} = q^{ab} \nabla_a n_b \,,
\end{equation}
respectively.  $\Surf$ is called a {\em trapped surface} if both
expansions are negative, a {\em marginally trapped surface} if
$\Theta_{(\ell)} = 0$ and $\Theta_{(n)} < 0$, and a {\em marginally
    outer trapped surface} (MOTS) if $\Theta_{(\ell)} = 0$ with no
condition on $\Theta_{(n)}$.  Note that there is still a freedom to
scale the null normals by some positive function $f > 0$, i.e.
$\ell^a \to f\ell^a$, $n^a \to f^{-1}n^a$, without affecting the cross
normalization.  This, however, leaves the signs of the two expansions,
and in particular the condition $\Theta_{(\ell)} = 0$, invariant.
Thus the above definitions do not depend on the scaling of the
null-normals.

A hypersurface $\HH$ will here be called a {\em Quasi-Local Horizon}
(QLH) if it admits a foliation of MOTSs $\Surf$.  Starting from a MOTS
on a Cauchy surface, QLHs arise by a time evolution.  The question of
whether the time evolution leads to a smooth QLH is studied in
\cite{Andersson:2005gq,Andersson:2007fh,Booth:2006bn}. It is shown
that the time evolution leads to a smooth $\HH$ if the MOTS satisfies
a stability condition
which turns out to be equivalent to the spectral properties of an
elliptic operator on $\Surf$.  The properties of $\HH$ have also been
studied elsewhere; see
\cite{Booth:2005qc,Ashtekar:2004cn,Ashtekar:2025wnu} for reviews.  The
focus of this paper will not be $\HH$ but rather the MOTSs themselves.

The work presented here is made possible by the high accuracy MOTS
finder which allows us to locate highly distorted MOTSs
\cite{pook_kolb_daniel_2019_2591105}.  This is an adaptation of a
previous method developed by Thornburg
\cite{Thornburg:2003sf,Thornburg:2006zb} which is restricted to
``star-shaped'' surfaces. If $\Surf$ is a star shaped surface,
then there exists a point $p$ such that every ray from $p$ intersects
$\Surf$ exactly once. Thus, every point on $\Surf$ can be
uniquely parametrized in terms of distance from $p$.  The method of
\cite{pook_kolb_daniel_2019_2591105} removes this restriction and
instead parametrizes $\mathcal{S}$ using distance from a
\emph{reference surface} which can itself be arbitrarily distorted.





\begin{figure*}
  \includegraphics[width=1\linewidth]{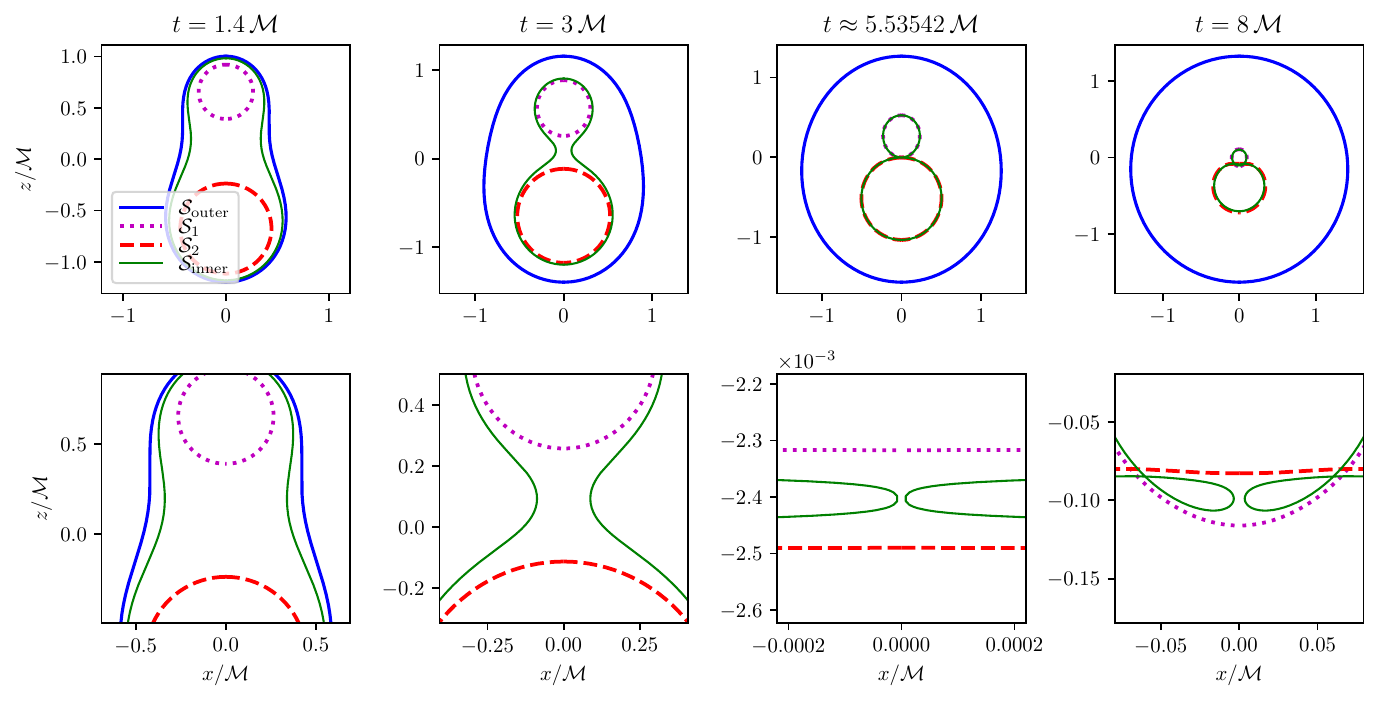}
  \caption{\label{fig:overview}%
    MOTS structure during the head-on collision.  The second row
    shows close-ups of the configuration in the first row.  The
    third column is very close before
    $\ttouch \approx 5.5378\,\mathcal{M}$.
    The blue curve is $\Sout$, $\Sin$ is in green, $\Sone$ is the
    dashed magenta curve, and finally $\Stwo$ is the dashed-red
    curve. }
\end{figure*}

With this MOTS finder at hand, it has been possible to understand how,
beginning with two distinct and widely separated MOTSs, we end up
eventually with a single MOTS corresponding to the remnant black hole.
When the two MOTSs get sufficiently close to each other, a common MOTS
appears which encloses the two progenitors BHs.  This common MOTS,
initially highly distorted, moves outwards and settles down to the
final remnant black hole, \emph{absorbing} gravitational radiation in
the process.  The recent progress made in
Refs.~\cite{PhysRevLett.123.171102,PhysRevD.100.084044} shows that a
connected sequence of MOTSs does in fact exist which connects the two
initial MOTSs to the final MOTS.  Let us summarize the general
behavior (see Fig.~\ref{fig:overview}):
In a head-on merger of two non-spinning black holes, the two
individual horizons $\Sone$ and $\Stwo$ were found to approach each
other until they touch (at a time denoted $\ttouch$) and subsequently
start to intersect.  A little before $\ttouch$, a common MOTS
surrounding $\Sone$ and $\Stwo$ appears which immediately bifurcates
into two, an inner and outer branch, denoted $\Sin$ and $\Sout$,
respectively.
We call the time of bifurcation $\tbifurcate$.
The outer branch $\Sout$ expands outwards and becomes
more and more symmetric as it approaches the event horizon of the
final Schwarzschild black hole.  The inner branch $\Sin$ on the other
hand, becomes increasingly distorted and at $\ttouch$, the union
$\Sone \cup \Stwo$ coincides with $\Sin$, with a cusp at the common
point.  It was found that $\Sin$ continues to evolve from this point
to the future, where it has a self-intersection.
From $\ttouch$ to the past, $\Sin$ evolves smoothly,
first to the past and then turning to the
future at $\tbifurcate$, to eventually asymptote to the final
Schwarzschild horizon. As
$\Sin$ turns to the future, we call it $\Sout$ and $(\Sout, \Sin)$ can
be seen as the two branches of the common horizon.  With the exception
of the MOTS at $\ttouch$, all other MOTSs shown here are seen to be
smooth.\footnote{%
  Self-intersecting MOTSs are smooth in the sense that they are
  smoothly immersed $2$-spheres.}  Several mathematical results are
known in the scenario described above; we mention one which will be
useful for us later.  At $\ttouch$, it turns out that $\Sone$ and
$\Stwo$ have the same mean curvature at the point of contact
\cite{Moesta:2015sga}.  Thus, at the point of contact, the ingoing
expansions of $\Sone$ and $\Stwo$ are identical.  Equivalently,
considering $\Sone$ and $\Stwo$ to be embedded in a Cauchy surface
$\Sigma$, the mean curvatures (i.e. the trace of the second
fundamental forms) will coincide at the point of contact.  It is also
shown in \cite{Andersson:2008up} that when $\Sone$ and $\Stwo$ get
sufficiently close, then they must be enclosed by a common MOTS.

Our goal here is to elucidate details of the cusp formation on $\Sin$,
and to also understand the geometry of $\Sone$ and $\Stwo$ in this
process.  This whole picture hinges on the fact that $\Sin$ tends to
$\Sone \cup \Stwo$ as $\tname \to \ttouch$ from both sides, and hence
necessarily develops a {\em cusp} during its evolution.  However, this
has not yet been fully understood.  First, consider $\Sin$ at
$\ttouch$ where it consists of $\Sone \cup \Stwo$ and has a cusp.
Given that the radius of curvature of $\Sin$ is necessarily zero at
the cusp, how does it {\em approach} zero and still be equal to the
radius of curvature of $\Sone$ and $\Stwo$, which clearly do {\em not}
go to zero?  In other words, will the radius of curvature jump
discontinuously at the cusp?  Furthermore, on either $\Sone$ or
$\Stwo$, how do their radii of curvature (and their angular
derivatives) behave near the point of touching?  How are these radii
of curvature related to each other while still maintaining the
Gauss-Bonnet theorem?  In this paper we will give a detailed
answer to these questions.

\subsection{Axisymmetric MOTSs and multipole moments}
\label{subsec:axisymmetry}

The main analytical tool we use here are suitable geometric multipole
moments of a distorted horizon.  Since the simulation considered in
the present paper is manifestly axisymmetric, we are able to simplify
the representation of axisymmetric surfaces.  Without loss of
generality, let the symmetry axis be the $z$ axis of our numerical
coordinates.  Any such surface will then be considered the surface of
revolution around the $z$ axis of a {\em curve} $\gamma$ in the
$(x,z)$ half plane.

On each axisymmetric MOTS $\Surf$, we will construct uniquely defined
coordinates $(\zeta,\varphi)$ using the procedure presented in
Ref.~\cite{Ashtekar:2004gp} which is briefly summarized here; this
construction applies in fact to any axisymmetric 2-surface.  Let
$\phi^a$ be the axial Killing vector on $\Surf$ and we assume that it
vanishes at exactly 2 points on $\Surf$, these are the poles on
$\Surf$, and also that it has closed integral curves. Let $A_\Surf$ be
the area of $\Surf$ and $R_\Surf = \sqrt{A_\Surf/4\pi}$ the area
radius.  The coordinate $\varphi$ is the affine parameter along the
integral curves of $\phi^a$, normalized so that
$0\leq \varphi < 2\pi$. The other coordinate $\zeta$ is defined via
\begin{equation}
  \label{eq:zeta-defn}
  \mathcal{D}_a\zeta = \frac{4\pi}{A_\Surf}\widetilde{\epsilon}_{ba}\varphi^b\,,\quad \oint_\Surf\zeta\,dA = 0\,.
\end{equation}
Here $D_a$ is the covariant derivative on $\Surf$ compatible with
$q_{ab}$, and $\epsilon_{ab}$ is the geometric volume 2-form on
$\Surf$.   The 2-metric on $\Surf$ has the simple form
\begin{equation}
  \label{eq:canonical-metric}
  ds^2_q = R_\Surf^2\left(\frac{d\zeta^2}{F(\zeta)} + F(\zeta)d\phi^2 \right) \,,
\end{equation}
where
\begin{equation}
  \label{eq:Fzeta-defn}
  F(\zeta) = \frac{4\pi\varphi_a\varphi^a}{A_\Surf}\,.
\end{equation}
The invariant coordinate $\zeta \in [-1, 1]$ thus becomes a parameter
along $\gamma$.  For the standard ``round'' 2-sphere in Euclidean
space with spherical coordinates $(\theta,\phi)$, it is
straightforward to verify that $F(\zeta) = 1-\zeta^2$ and
$\zeta = \cos\theta$.  We shall refer to $F$ on the ``round'' 2-sphere
as $F_o = 1-\zeta^2$.  We will often define $\theta = \cos^{-1}\zeta$
in this way even on distorted spheres and show results as functions of
$\theta$; we shall usually suppress the dependence on the orthogonal
angular coordinate $\varphi$.  Note that we could in principle choose
a different range of values for $\zeta$ such that
$\zeta_{min} \leq \zeta \leq \zeta_{max}$.  However, the total area
will still be
$2\pi R_\Surf^2(\zeta_{max}-\zeta_{min}) = 4\pi R_{\Surf}^2$.  It
follows that $\zeta_{max}-\zeta_{min} = 2$, and it is convenient to
choose $\zeta_{max}=1$ and $\zeta_{min}=-1$. The point $\zeta = +1$ is
referred to as the north pole, and $\zeta=-1$ as the south pole.

In several places, it will be useful to parametrize the curve $\gamma$ using
its normalized proper length parameter, which is related to $\zeta$ via
\begin{equation}\label{eq:stilde}
    \stilde(\zeta) := \frac{\int_{\zeta}^1 F(\zeta')^{-1/2}\,d\zeta'}{\int_{-1}^1 F(\zeta')^{-1/2}\,d\zeta'}
    \,.
\end{equation}
This is zero at the north pole and one at the south pole.

A few details about the canonical metric
Eq.~(\ref{eq:canonical-metric}) will be useful for us later. First,
since the axial Killing vector $\phi^a$ vanishes at the poles, we see
from Eq.~(\ref{eq:Fzeta-defn}) that
\begin{equation}
  \label{eq:Flimit}
  \lim_{\zeta \rightarrow \pm 1} F(\zeta) = 0\,.
\end{equation}
Second, in order to avoid conical singularities at the pole, we must
have
\begin{equation}
  \label{eq:Fplimit}
  \lim_{\zeta \rightarrow \pm 1}F^\prime(\zeta) = \mp 2
\end{equation}
These conditions also hold if $\zeta_{max}$ and $\zeta_{min}$ differ
from $+1$ and $-1$ respectively (as long as $F$ vanishes at these
points).

The scalar curvature $\RR$ calculated from the
metric Eq.~(\ref{eq:canonical-metric}) is
\begin{equation}
  \label{eq:RRF}
  \RR(\zeta) = -\frac{1}{R_\Surf^2}F^{\prime\prime}(\zeta)\,.
\end{equation}
Note that the geometric volume element corresponding to the metric
Eq.~(\ref{eq:canonical-metric}) is $d^2V = R_\Surf^2 d\zeta\,d\phi$ and
is thus independent of $F$, and therefore the same as on a ``round''
2-sphere.  This fact is important when discussing the orthogonality of
spherical harmonics. We can define the spherical harmonics
$Y_{\ell m}(\theta,\phi)$ (with $\theta = \cos^{-1}\zeta)$ on any
distorted axisymmetric sphere.  These would then satisfy the same
orthogonality relationship with the geometric volume element on
$\Surf$ as for the ``round'' 2-sphere.  We will follow the
normalization
\begin{equation}
  \oint_\Surf Y^m_{\ell}Y^{m^\prime\star}_{\ell^\prime} d^2V = R_\Surf^2 \delta_{\ell\ell^\prime}\delta_{mm^\prime}\,.
\end{equation}
In this paper we will use the $m=0$ harmonics $Y_{\ell 0}$. In terms
of the Legendre polynomials $P_{\ell}(\zeta)$, these are
\begin{equation}
  Y^0_{\ell}(\theta,\phi) = \sqrt{\frac{(2\ell+1)}{4\pi}}P_\ell(\zeta)\,.
\end{equation}
The above coordinate system can be employed to construct invariant
geometric multipole moments on an axisymmetric MOTS. For a non-spinning
axisymmetric QLH, the key geometric quantity of interest is the
intrinsic scalar curvature $\RR$ on a MOTS. First, we recall the well
known Gauss-Bonnet theorem relating $\RR$ on a compact 2-manifold
$\Surf$ (with volume element $d^2V$) without boundary to its genus
$g$:
\begin{equation}
  \oint_\Surf \RR\,d^2V = 8\pi (1-g)\,.
\end{equation}
For a sphere $g=0$ so that
\begin{equation}\label{eq:gauss-bonnet}
  \oint_\Surf \RR\,d^2V = 8\pi\,.
\end{equation}
Here we shall deal only with topological spheres; even the MOTSs
with self-intersections are, topologically speaking, spheres.  In the
invariant coordinate system, it is easy to verify the Gauss-Bonnet
theorem by an explicit integration using Eqs.~(\ref{eq:RRF}) and
(\ref{eq:Fplimit}).

We define then geometric multipole moments on $\Surf$ by decomposing
$\RR$ using the spherical harmonics
\cite{Ashtekar:2004gp,Ashtekar:2013qta}
\begin{eqnarray}
  \label{eq:Ilexplicit}
  I_\ell &=& \frac{1}{4}\oint_\Surf \RR Y_{\ell 0}(\zeta,\phi)\,d^2V\\
  &=& \pi R_\Surf^2\sqrt{\frac{(2\ell+1)}{16\pi}}\int_{-1}^1 \RR(\zeta) P_\ell(\zeta)\,d\zeta \\
  &=& -\pi\sqrt{\frac{(2\ell+1)}{16\pi}}\int_{-1}^1 F^{\prime\prime}(\zeta) P_\ell(\zeta)d\zeta\,.
\end{eqnarray}
In principle, if we were to use different angular coordinates on
$\Surf$, the values of the multipole moments would differ. Using the
geometrically defined $(\zeta,\phi)$ removes this ambiguity.  Also,
since we shall only consider axisymmetric surfaces, $\RR$ will have no
$\phi$ dependence and we do not need to consider
$Y_{\ell m}(\zeta,\phi)$ (and thus $I_{\ell m}$) with non-vanishing
$m$.  Using again the Gauss-Bonnet theorem, we see that $I_0=\sqrt{\pi}$.
Finally, from Eq.~(\ref{eq:Ilexplicit}) it can be shown that $I_1=0$
since
\begin{eqnarray}
  I_1 &=& -\sqrt{\frac{3\pi}{16}}\int_{-1}^1 F^{\prime\prime}(\zeta)\zeta \,d\zeta \\
  &=& -\sqrt{\frac{3\pi}{16}}\left\{\left.(F^\prime\zeta - F)\right|_{-1}^1   \right\} = 0\,.\label{eq:I1Vanish}
\end{eqnarray}
We have used Eqs.~(\ref{eq:Flimit}) and (\ref{eq:Fplimit}).

Since $I_1$ vanishes identically, we need only consider $\ell\geq 2$.
The above \emph{geometric} multipoles are dimensionless, and must be
rescaled to get the physical horizon mass multipole moments:
\begin{equation}
  \scrM_\ell = \sqrt{\frac{4\pi}{2\ell + 1}}\frac{MR_\Surf^\ell}{2\pi}I_\ell\,.
\end{equation}
Here $M = \frac{R_\Surf}{2}$ is the horizon mass (for a non-spinning
BH).  In this case
\begin{equation}
  \scrM_\ell = \sqrt{\frac{4\pi}{2\ell + 1}}\frac{R_\Surf^{\ell+1}}{4\pi}I_\ell\,.
\end{equation}
The vanishing of the mass dipole $\scrM_1$ can be taken to imply that
we are in the center of mass frame.  There are also appropriate
definitions for angular momentum and spin multipole moments, but we
shall not focus on them in this paper.  See \cite{Gourgoulhon:2026nes}
for a comparison of different notions of multipole moments for a Kerr
black hole.

\subsection{The numerical set-up}

The initial data is time-symmetric Brill-Lindquist data
\cite{Brill:1963yv} for two non-spinning and uncharged black holes.
Thus, we consider Euclidean space $\mathbb{R}^3$ with two points (the
``punctures'') $\mathbf{x}_{1,2}$ excluded.  Let $\delta_{ij}$ be the
Euclidean metric on $\mathbb{R}^3$, and $d$ the Euclidean distance
between the punctures: $d=|\mathbf{x}_1-\mathbf{x}_2|$.  The extrinsic
curvature is trivial: $K_{ij}=0$, and the 3-metric is conformally flat:
$h_{ij} = \Phi^4\delta_{ij}$.  The conformal factor which satisfies
the constraint equations of vacuum general relativity, and
$\Phi\rightarrow 1$ at spatial infinity, is
\begin{equation}
  \Phi(\mathbf{x}) = 1 + \frac{m_1}{2\left|\mathbf{x} - \mathbf{x}_1\right|} + \frac{m_2}{2\left|\mathbf{x} - \mathbf{x}_2\right|}\,.
\end{equation}
Despite the simplicity, this data contains a rather complicated set of
MOTSs.  For a sequence of initial data with $m_{1,2}$ fixed and $d$
decreasing, it is seen that for large $d$, the initial data contains
two separate MOTSs corresponding to the two progenitor BHs.  As $d$ is
decreased, a common horizon appears at a particular value of $d$ which
in turn bifurcates into an inner and outer branch, rather similar to
what is seen during time evolution; see \cite{Pook-Kolb:2018igu} for
details.

As mentioned earlier, we use the same simulation as in
Refs.~\cite{PhysRevLett.123.171102,PhysRevD.100.084044}. The bare
masses are $m_{1} = 0.5$ and $m_2 = 0.8$.
The total mass of our simulation is hence $m_1 + m_2 = 1.3$ and our plots and
numerical values use this mass, which we denote with $\mathcal{M}$.
The punctures are located
on the $z$ axis, initially at $\pm 0.65\,\mathcal{M}$.  The exact parameter files
are available in the repository \cite{pook_kolb_daniel_2019_3352328}.
No mesh refinement or multiple grid resolutions are used.  The simulations
are performed with the \emph{Einstein Toolkit} \cite{Loffler:2011ay,
  EinsteinToolkit:web} and \emph{TwoPunctures} \cite{Ansorg:2004ds}
for the initial conditions, and \emph{McLachlan} and \emph{Kranc}
\cite{Brown:2008sb,Husa:2004ip, Kranc:web} to solve the Einstein
equations.

\section{The area and mass at the merger}
\label{sec:ricci}

\begin{figure*}
  \centering
  \includegraphics[width=.45\linewidth]{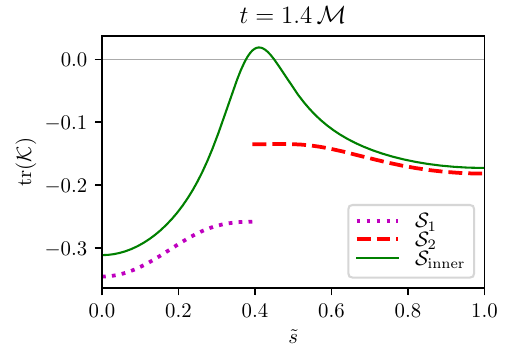}\hfill
  \includegraphics[width=.45\linewidth]{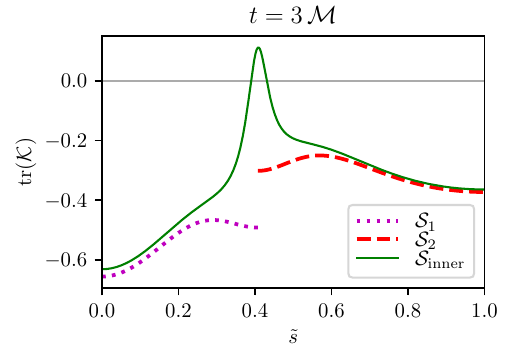}\\
  \includegraphics[width=.45\linewidth]{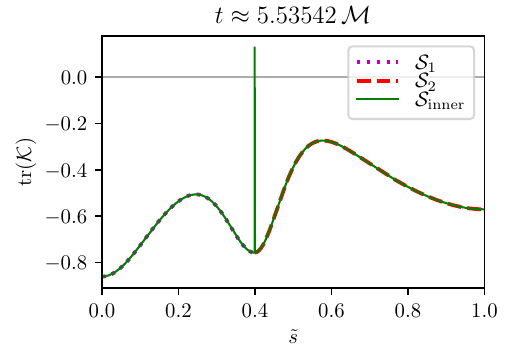}\hfill
  \includegraphics[width=.45\linewidth]{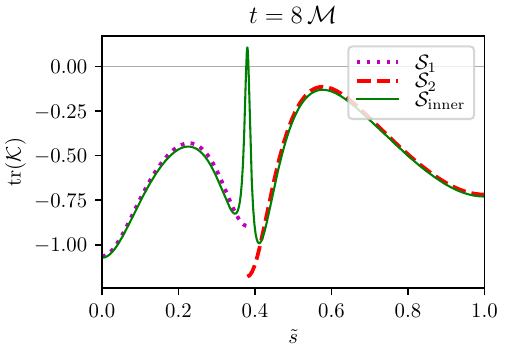}%
  \caption{\label{fig:mean_curv_compare}%
    Mean curvature $\mathop{\mathrm{tr}}(\mathcal{K})$ of $\Sin$
    (green) and the two individual MOTSs (dashed purple and red). The
    parameter $\stilde$ on the x-axis here is the proper length of a
    constant-$\phi$ curve on $\Sone$, $\Stwo$; these parameters on
    $\Sone$ and $\Stwo$ have been transformed linearly so that they
    respectively end or start at the location $\sneck$ (where $\Sin$
    has a locally minimal circumference away from its poles). This
    allows us to compare the various mean curvatures on the same plot
    even away from $\ttouch$ where $\Sone$, $\Stwo$ and $\Sin$ are
    distinct surfaces. }
\end{figure*}

With the necessary background and the tools outlined above, we are now
ready to start studying the cusp at the merger. There are three MOTS
sequences of interest to us: $\Sin$, $\Sone$ and $\Stwo$.  The
behavior of $\RR$ on $\Sone$ and $\Stwo$ is relatively
straightforward: it remains smooth through the merger even when they
touch and subsequently intersect each other. On the other hand, $\Sin$
is more complicated since it develops a narrow ``neck'' leading to a
cusp at the time $\ttouch$.  $\Rin$ takes increasingly large negative
values at the neck and then becomes singular at the cusp exactly at
$\ttouch$. Subsequently, when $\Sin$ develops self-intersections,
$\RR$ is again smooth.  Therefore, we will focus here on the behavior
of $\Rin$ near $\ttouch$.  Consistent with the scenario sketched in
Sec. \ref{subsec:qlh}, the mean curvatures of $\Sone$ and $\Stwo$
coincide at the point of contact; see
Fig.~\ref{fig:mean_curv_compare}. The mean curvature of $\Sin$ is seen
to have a spike around the neck and around $\ttouch$; the region with
positive mean curvature shrinks near $\ttouch$.

At $\ttouch$ we would like to relate the masses of $\Sone$ and $\Stwo$
with the mass of $\Sin$.  This is straightforward because at
$\ttouch$, we must have
\begin{equation}
  A_{in} = A_1+A_2\,,
\end{equation}
which implies, $R_{in}^2 = R_1^2 + R_2^2$.  Thus, since these are
non-spinning black holes, at $\ttouch$, the mass of $\Sin$ is obtained
from the masses of $\Sone$ and $\Stwo$ as
\begin{equation}
  M_{in} = \sqrt{M_1^2 + M_2^2}\,.
\end{equation}
This value of $M_{in}$ can then be used
as an initial value as we follow $\Sin$, initially backwards in time,
through $\tbifurcate$ and then eventually towards the final black
hole.  It follows that $M_{in} < M_1 + M_2$ and moreover
\begin{eqnarray}
  M_{in} &=& (M_1+M_2)\sqrt{1-\frac{2M_1M_2}{(M_1+M_2)^2}}\nonumber \\
  &=& (M_1+M_2)\sqrt{1-2\eta}
\end{eqnarray}
with $\eta = M_1M_2/(M_1+M_2)^2$ being the symmetric mass ratio.
Since $0\leq \eta \leq 0.25$, we must have $M_{in} \leq M_1+M_2$.  For
equal masses, $\eta=0.25$ which leads to the largest value of
$M_1+M_2-M_{in} = (M_1+M_2)(1-1/\sqrt{2})$ (this is the same limit as
obtained in \cite{PhysRevLett.26.1344}). For highly asymmetric
systems, i.e. small $\eta$, we will have $M_{in} \approx M_1+M_2$.
When we connect this to the final remnant mass (by tracing $\Sin$
\emph{backwards} in time till we reach $\tbifurcate$, and then forward
in time till we reach the final state of the remnant) the final mass
will be larger than the above value of $M_{in}$.  In greater detail,
if we trace the mass of the system using the horizon masses of the
above sequence of MOTSs, we get three distinct effects:
\begin{enumerate}
  \item The behavior of $\Sone$ and $\Stwo$: the mass is $M_1+M_2$ which
        increases monotonically for $-\infty<t<\ttouch$, as the two horizons
        absorb infalling radiation.
  \item At $\ttouch$: A discontinuous decrease in the mass when we go
        from $M_1+M_2 \rightarrow M_{in}$.
  \item Approach to equilibrium: $M_{in}$ generally increases till we
        reach the final equilibrium state (this increase is however not
        always monotonic \cite{PhysRevLett.123.171102}).
\end{enumerate}
The discontinuous jump in (2) above is larger (in absolute value) than
the mass increases in (1) and (3), so that the final remnant mass is
less than the sum of the initial masses.  This is consistent with
the corresponding scenario at null infinity, where the flux of
gravitational waves carries away energy from the system leading to a
mass-loss.

\section{The geometry of $\Sin$ near $\ttouch$}
\label{sec:nearttouch}

Going beyond the mass to the higher multipole moments requires a more
detailed analysis of the geometry of $\Sin$ (and its relation to
$\Sone,\Stwo$) near $\ttouch$.  Since $\Sin$ (like the other MOTSs) is
axisymmetric, we can use the invariant coordinates defined earlier.
Let $(\zeta,\phi)$ (with $-1\leq \zeta < 1$ and $0\leq \phi< 2\pi$) be
the invariant coordinate system on $\Sin$ and let $\RR(\zeta)$ be the
scalar curvature; it is independent of $\phi$ due to axisymmetry.
Define the angle $\theta = \cos^{-1}\zeta$. Let $\zetaneck$ be the
location of the neck, i.e. the narrowest point of $\Sin$.
Let further $\thneck$ and $\sneck$ be the corresponding values of
$\theta$ and $\stilde$.
The
circumference of a constant-$\zeta$ curve is
seen to be $2\pi R\sqrt{F(\zeta)}$, whence the narrowest point
$\thneck$ corresponds to a minimum of $F(\zeta)$.  When considering
the behavior of the Ricci scalar $\Rin$ (and $F$) of $\Sin$, we need
to analyze two limits, the one of $\theta\to\thneck$ and the limit
$\tname\to\ttouch$.  Fixing a $\theta \neq \thneck$, the Ricci scalar
converges to that of $\Sone$ or $\Stwo$ (depending on whether
$\theta < \thneck$ or $\theta > \thneck$).

\begin{figure*}
  \includegraphics[width=.45\linewidth]{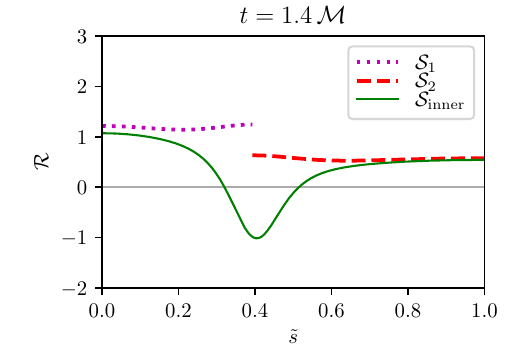}\hfill
  \includegraphics[width=.45\linewidth]{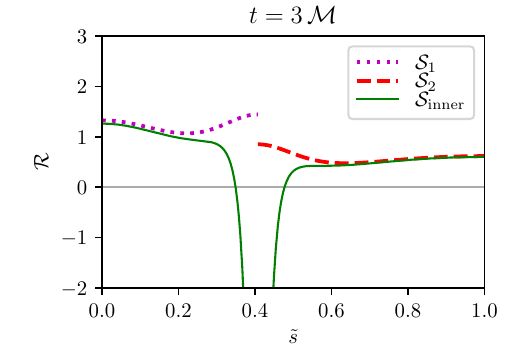}\\
  \includegraphics[width=.45\linewidth]{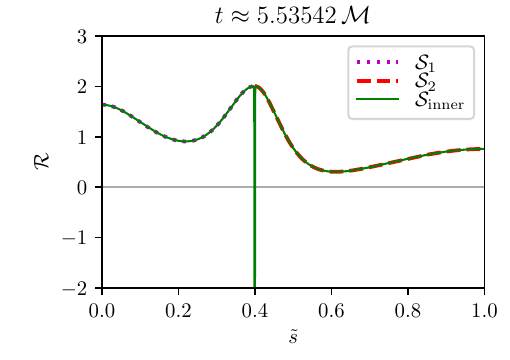}\hfill
  \includegraphics[width=.45\linewidth]{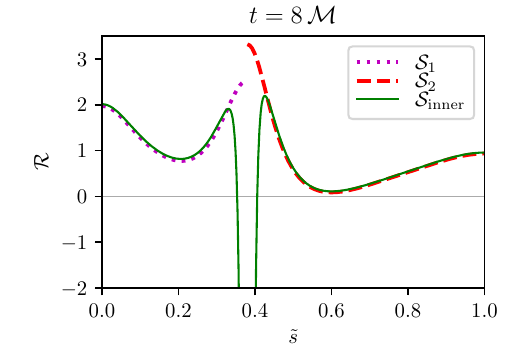}%
  \caption{\label{fig:ricci_compare}%
    Ricci scalar $\mathcal{R}$ of $\Sin$ and the two individual
    MOTSs. The curve parameter of $\Sone$, $\Stwo$ has been
    transformed linearly to end or start at
    $\sneck$, respectively.
  }
\end{figure*}

Fig.~\ref{fig:ricci_compare} shows the Ricci scalars of $\Sone$,
$\Stwo$ and $\Sin$ with the curve parameters suitably rescaled and
shifted so that the values can be compared with $\Sin$; for $\Sone$ we
will have $0<\theta<\thneck$, while for $\Stwo$ we will have
$\thneck<\theta<\pi$. We see that $\Rone$ and $\Rtwo$ are
everywhere positive while $\Rin$ is negative around the neck.
Moreover, from the bottom left panel of Fig.~\ref{fig:ricci_compare}
it is evident that the curvatures on $\Sone$ and $\Stwo$ agree at the
point of contact.  This issue, along with the question of the
coordinate singularity at the poles, is discussed further in Appendix
\ref{app:ricci} and in Appendix \ref{app:ricci_derivatives}.

\begin{figure*}
  \includegraphics[width=.45\linewidth]{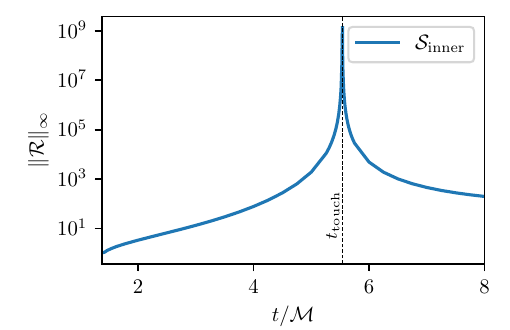}\hfill
  \includegraphics[width=.45\linewidth]{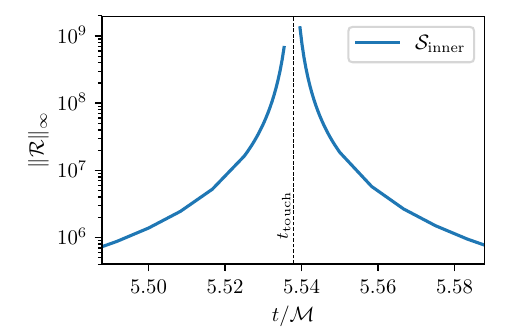}%
  \caption{\label{fig:ricci_max_inner}%
    Maximum value of the Ricci scalar $|\mathcal{R}|$ along $\Sin$ as
    function of time. The right panel shows a close-up near $\ttouch$.
    We find numerical indication that the Ricci scalar diverges as
    $\tname\to\ttouch$.
  }
\end{figure*}

Let $\RR_{12} : [0, \pi]\setminus\{\thneck\} \to \mathbb{R}$ be the
values of the Ricci scalar $\RR$ along first $\Sone$ and then $\Stwo$,
with the transition happening at the neck of $\Sin$, i.e.
\begin{equation}\label{eq:R12}
  \RR_{12}(\theta) := \begin{cases}
    \displaystyle\RR_1\left(\pi\,\frac{\theta}{\thneck}\right)\,,
     & 0 \leq \theta < \thneck   \\[2ex]
    \displaystyle\RR_2\left(\pi\,\frac{\theta-\thneck}{\pi-\thneck}\right)\,,
     & \thneck < \theta \leq \pi
  \end{cases}\,.
\end{equation}
We see that at $\ttouch$, if the negative spike were to be
excluded, the resulting function for $\Rin$ would be everywhere
continuous and differentiable, and would in fact agree with
$\RR_{12}$.  However, $\RR_{12}$ cannot, by itself, be the correct
model for $\Rin$ since it would give the incorrect
topology as far as the Gauss-Bonnet theorem is concerned.
At the neck, the plot of the maximum of $|\Rin|$ shown in
Fig.~\ref{fig:ricci_max_inner} indicates that $\mathcal{R}$ diverges to
$-\infty$ as $\tname\to\ttouch$.  Thus, at $\ttouch$, the difference
between $\Rin$ and $\RR_{12}$ is a $\delta$-function normalized
precisely so that the Gauss-Bonnet formula is satisfied



\begin{figure*}
  \includegraphics[width=0.45\linewidth]{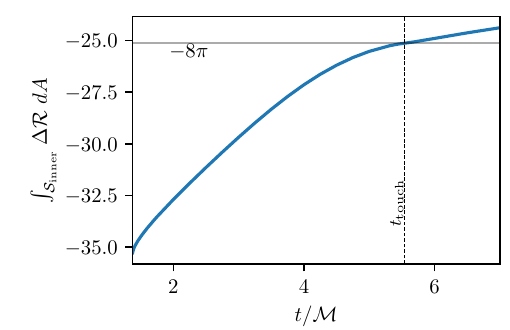}\hfill
  \includegraphics[width=0.45\linewidth]{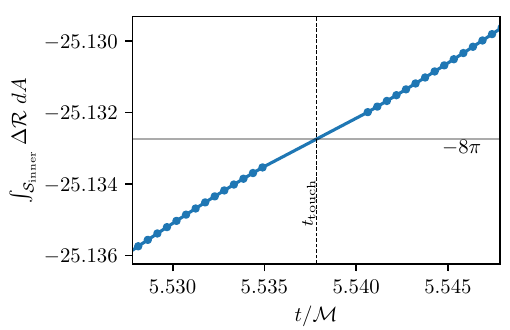}%
  \caption{\label{fig:DeltaR_integral}%
    Integral of $\Delta\RR$ as function of simulation time.
    This shows our numerical support for Eq.~\eqref{eq:delta_ricci_integral}.
  }
\end{figure*}

\begin{figure*}
  \includegraphics[width=0.45\linewidth]{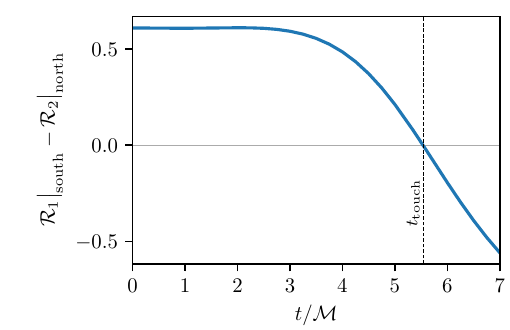}\hfill
  \includegraphics[width=0.45\linewidth]{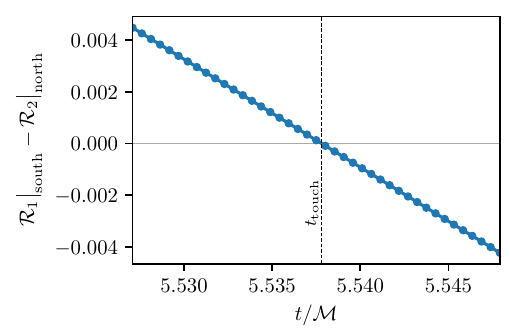}%
  \caption{\label{fig:S12_ricci_delta_plots}%
    Difference of the Ricci scalar at the south pole of $\Sone$ and north
    pole of $\Stwo$. This shows that the Ricci scalars coincide at the two
    points that touch at $\ttouch$.
  }
\end{figure*}

If we write $\Rin = \RR_{12}+\Delta\RR$, then
Eq.~\eqref{eq:gauss-bonnet} implies
\begin{equation}\label{eq:delta_ricci_integral}
  \lim_{\tname\to\ttouch}\int_{\Sin} \Delta \RR\ dA = -8\pi\,,
\end{equation}
provided $\Sin$ merges with $\Sonetwo$ at $\ttouch$.  The above
integral is shown in Fig.~\ref{fig:DeltaR_integral} at times around
$\ttouch$.  This shows that we have strong numerical support for the
validity of Eq.~\eqref{eq:delta_ricci_integral}.  Additionally, we
have mentioned earlier that the facing points on the poles of
$\Sonetwo$ take on the same value of the Ricci scalar at exactly
$\ttouch$, which is the time when these two points meet.
Fig.~\ref{fig:ricci_compare} already suggests this possibility and
Fig.~\ref{fig:S12_ricci_delta_plots} shows that we indeed find a
strong numerical indication for this fact.

It is also instructive to see what the above implies for the behavior
of $F$ at $\ttouch$.  Since $\Rin$ is the second derivative of $F$, the
$\delta$-function singularity of $\Rin$ at $\zetaneck$ implies a
discontinuity of $F^\prime$ at $\zetaneck$.  At $\ttouch$, we thus
deduce the following facts about $F$ near $\zetaneck$.  The neck has
vanishing circumference:
\begin{equation}
  \lim_{\zeta\rightarrow\zetaneck}F(\zeta) = 0\,.
\end{equation}
Note that $F$ is positive-definite away from the poles and the neck.
The absence of conical singularities implies (as in
Eq.~\eqref{eq:Fplimit}) that
\begin{equation}\label{eq:limitsF}
  \lim_{\zeta\rightarrow{\zetaneck^{-}}}F^\prime(\zeta) = -2\,,\quad   \lim_{\zeta\rightarrow{\zetaneck^{+}}}F^\prime(\zeta) = 2\,.
\end{equation}
Thus, near the neck, $F^{\prime\prime}$ has a $\delta$-function
singularity: $F^{\prime\prime}\approx 4\delta(\zeta-\zetaneck)$.
Using the proper area element on $\Sin$, this is seen to be consistent
with Eq.~\eqref{eq:delta_ricci_integral} as it should be.  It is thus
natural to say that the neck of
$\Sin$ pinches off into the ``north-pole'' of
$\Stwo$ ($\zeta\rightarrow
  \zetaneck^{-}$) and the ``south-pole'' of $\Sone$ ($\zeta\rightarrow
  \zetaneck^{+}$).

Finally, the value of $\zetaneck$ at $\ttouch$ can be calculated as
follows.  Since $\Sin=\Sone\bigcup\Stwo$, the total area of $\Sin$
must equal the sum of the areas of $\Sone$ and $\Stwo$.  The area of
$\Sin$ from the south-pole $\zeta_{in}=-1$ up to $\zeta=\zetaneck$ is
$2\pi R_{in}^2(\zetaneck+1)$.  But this should also be equal to the
area of $\Sone$, i.e. $4\pi R_1^2$. Thus:
\begin{eqnarray}
  \label{eq:zetaneck}
  \zetaneck &=& \frac{2R_1^2}{R_{in}^2} -1 = 2\frac{M_1^2}{M_{in}^2} - 1\nonumber \\
  &=& \frac{M_1^2-M_2^2}{M_1^2+ M_2^2}\,.
\end{eqnarray}

\section{The mass multipole moments at the merger}
\label{sec:multipoles}

In this section we present analytical results regarding the mass
multipole moments of $\Sin$ at the merger, i.e. at $t=\ttouch$.

\subsection{The mass dipole moment at the merger}

The mass dipole moment $\scrM_1$ (or alternatively the geometric moment
$I_1$) of $\Sin$ has some features of interest as we shall now
discuss.  As mentioned earlier, $\scrM_1$ vanishes identically. Following
\cite{Ashtekar:2004gp}, this is interpreted to say that the
coordinates $(\zeta,\phi)$ place us in the center of mass of the
horizon.  Previously, we proved $I_1=0$ in
Eq.~\eqref{eq:I1Vanish}. However that derivation assumed that $F$
vanishes only at the poles, but now it vanishes additionally at
$\zetaneck$ and we must therefore revisit the vanishing of $I_1$ for
$\Sin$ at $\ttouch$.

Let us break up the integral for $I_1$ at the neck:
\begin{eqnarray}
  I_1 &=& -\sqrt{\frac{3\pi}{16}}\int_{-1}^1 F^{\prime\prime}(\zeta)\zeta \,d\zeta \nonumber\\
  &=& -\sqrt{\frac{3\pi}{16}}\left\{ \int_{-1}^{\zetaneck} F^{\prime\prime}(\zeta)\zeta \,d\zeta + \int_{\zetaneck}^1 F^{\prime\prime}(\zeta)\zeta \,d\zeta \right\}
\end{eqnarray}
At first glance, one might interpret the two individual integrals
appearing here with the dipole moments of $\Sone$ and $\Stwo$.  This
is however not correct and these two terms do not individually
vanish\footnote{In fact, we have here three different coordinate
  systems. On $\Sin$ we have the coordinates $(\zeta,\phi)$.  These
  are however not the same as the coordinates on $\Sone,\Stwo$.  If
  $(\zeta_1,\phi)$ and $(\zeta_2,\phi)$ are coordinates on $\Sone$ and
  $\Stwo$ respectively, then the mass dipole moments vanish in these
  coordinates.  However, this does not hold in the coordinates
  $(\zeta,\phi)$.}.  A simple calculation yields (using
Eq.~\ref{eq:limitsF}):
\begin{eqnarray}
  \int_{-1}^{\zetaneck} F^{\prime\prime}(\zeta)\zeta \,d\zeta  &=& \left.(F^\prime\zeta - F)\right|_{-1}^{\zetaneck} = 2-2\zetaneck\,,\\
  \int_{\zetaneck}^{1} F^{\prime\prime}(\zeta)\zeta \,d\zeta  &=& \left.(F^\prime\zeta - F)\right|_{-1}^{\zetaneck} = -2-2\zetaneck\,.
\end{eqnarray}
The sum therefore does not vanish:
\begin{equation}
  I_1 = \sqrt{\frac{3\pi}{16}}\times 4\zetaneck = \sqrt{3\pi}\frac{M_1^2-M_2^2}{M_1^2+ M_2^2}\,.
\end{equation}
We conclude that the mass-dipole moment of $\Sin$ does \emph{not}
vanish at $\ttouch$.  Immediately away from $\ttouch$ however $I_1$
does vanish.  We can interpret this behavior as a jump in the center
of mass corresponding to the value
\begin{equation}
  \Delta I_1 = \sqrt{3\pi}\frac{M_1^2-M_2^2}{M_1^2+ M_2^2}\,.
\end{equation}
This shift vanishes for an equal mass binary, and is otherwise in the
direction of the heavier black hole; it approaches $\pm \sqrt{3\pi}$
in the limit of extreme mass ratios.  The same results can be obtained
also directly from the scalar curvature.  Modeling $\Rin$ as the sum
of a smooth part and a $\delta$-function (as we shall discuss
shortly), the $\delta$-function is seen to be responsible for the
above shift in the center of mass.

\subsection{The mass quadrupole and higher moments}
\label{subsec:highermoments}

Let us now turn to the geometric multipole moments $I_\ell$ for
$\ell\geq 2$.  It will be easier to work directly with the scalar
curvature on $\Sin$ with its $\delta$-function singularity. Thus, we
take at $\ttouch$
\begin{equation}
  \Rin = \RR_{12}-\frac{4}{R_{in}^2}\delta(\zeta-\zetaneck) \,.
\end{equation}
Here $\RR_{12}$ is a smooth function whose integral over $\Sin$ is
$16\pi$; it coincides with $\Rone$ on the portion $\Sone$ and with
$\Rtwo$ on $\Stwo$.  Applying Eq.~(\ref{eq:Ilexplicit}) then leads to
three contributions for $\Sin$:
\begin{equation}
  I^{(in)}_\ell = I_\ell^{(in,1)} +  I_\ell^{(in,2)} +  I_\ell^{(cusp)}
\end{equation}
where
\begin{eqnarray}
  I_\ell^{(in,1)} &=& \pi R_{in}^2C_\ell\int_{-1}^{\zetaneck} \RR_1(\zeta) P_\ell(\zeta)d\zeta \,,\\
  I_\ell^{(in,2)} &=& \pi R_{in}^2C_\ell\int_{\zetaneck}^1 \RR_2(\zeta) P_\ell(\zeta)d\zeta \,,\\
  I_\ell^{(cusp)} &=& -8\pi^2 C_\ell\int_{-1}^1  \delta(\zeta-\zetaneck)P_\ell(\zeta)d\zeta \,.
\end{eqnarray}
We have defined
\begin{equation}
  C_\ell = \sqrt{\frac{(2\ell+1)}{16\pi}}\,.
\end{equation}
At $\ttouch$ we obtain
\begin{equation}
  I_{\ell}^{(cusp)} =  -8\pi^2\sqrt{\frac{(2\ell+1)}{16\pi}}P_\ell(\zetaneck)
\end{equation}
We see that the contribution of the cusp will generally be
non-vanishing unless $\zetaneck$ happens to lie at one of the zeros of
$P_\ell(\zeta)$.  As we have seen in Eq.~(\ref{eq:zetaneck}), the
value of $\zetaneck$ itself depends on the relative size of the two
black holes, and thus eventually on the mass ratio $q=M_2/M_1$ of the
binary system.  Thus, there should be values of $q$, depending on
$\ell$, for which this contribution should vanish at the cusp.  As an
example, consider the quadrupole $\ell=2$.  The zeros of $P_2(\zeta)$
occur at $\zeta=\pm(1/\sqrt{3})$. Taking then $\zetaneck=1/\sqrt{3}$
leads to $q^2 = (\sqrt{3}-1)/(\sqrt{3}+1)$, i.e. $q\approx 0.268$.
Thus, for configurations with mass ratios close to this value, we
should find correspondingly smaller deviations in the quadrupole
moment across the merger.  Similar considerations apply to the higher
multipoles.


Consider now the other contributions to the multipole moments due to
$\Rinone$ and $\Rintwo$ at $\ttouch$.  As before, we define $\Rinone$,
$\Rintwo$ as having support on $\zeta \in [-1, \zetaneck)$ and
$(\zetaneck, 1]$, respectively, such that away from $\zetaneck$,
$\Rinone + \Rintwo = \RR_{12}$.  However, at $\ttouch$ itself, the two
will differ by the $\delta$-function at the cusp.  Since
$\Sone\bigcup\Stwo = \Sin$ at $\ttouch$, we must be careful to
distinguish between the different $\zeta$ coordinates on $\Sone$,
$\Stwo$ and $\Sin$.  On these surfaces, we will have coordinates
$\zeta_1,\zeta_2,\zeta_{in}$ each of which takes values within the
range $[-1,1]$. Thus, even though $\Rinone$ agrees exactly with
$\Rone$, we need to consider that $\Rone(\zeta_1)$ is defined for
$-1\leq \zeta_1 \leq 1$ while for $\Rinone(\zeta_{in})$ we will have
$-1\leq \zeta_{in} < \zetaneck$.  Note that
\begin{equation}
  \int_{-1}^1\RR_{12} P_\ell d\zeta = \int_{-1}^{\zetaneck}\Rinone P_\ell d\zeta + \int_{\zetaneck}^1\Rintwo P_\ell d\zeta
  \,.
\end{equation}
At $\ttouch$, each of the terms on the right-hand-side should be
closely related to the multipole moments on $\Sone$ and $\Stwo$
respectively. For clarity, at $\ttouch$, we shall distinguish between
$\zeta_{in}$, $\zeta_{1}$ and $\zeta_2$. Thus, for example, by
changing variables from $\zeta_{in}$ to $\zeta_1$, the first term becomes
\begin{eqnarray}
  &&\!\!\!\!\!\!\!\!\int_{-1}^{\zetaneck}\Rinone(\zeta_{in}) P_\ell(\zeta_{in}) d\zeta_{in} \nonumber\\
  &=& \int_{-1}^1\Rinone(\zeta_{in}(\zeta_1)) P_\ell(\zeta_{in}(\zeta_1)) \frac{d\zeta_{in}}{d\zeta_1} d\zeta_1 \nonumber\\
  &=& \int_{-1}^1\Rone(\zeta_1) P_\ell(A\zeta_1 + B) \frac{d\zeta_{in}}{d\zeta_1} d\zeta_1 \nonumber\\
  &=& A\int_{-1}^1\Rone(\zeta_1) P_\ell(A\zeta_1 + B)d\zeta_1 \,.
\end{eqnarray}
In the third line we have assumed that (at $\ttouch$), at points of
$\Sone$, $\Rinone$ is identical to the scalar curvature $\Rone$.  To
obtain the last line, note that the transformation $\zeta_{in}(\zeta_1)$
is linear
\begin{equation}
  \zeta_{in} = A\zeta_1 + B
\end{equation}
where
\begin{eqnarray}
  A &=& \frac{M_2^2}{M_1^2+ M_2^2} = \frac{1}{1+q^2}\,,\\
  B &=& -\frac{M_1^2}{M_1^2+ M_2^2}= -\frac{q^2}{1+q^2}\,.
\end{eqnarray}
This transformation gives the expected results: $\zeta_{in}(-1) = -1$
and $\zeta_{in}(1) = \zetaneck$ consistent with
Eq.~(\ref{eq:zetaneck}). Thus, we have the following result: The
multipole moments of $\Rinone(\zeta_{in})$ viewed as the curvature of
$\Sin$, are \emph{not} the same as the multipole moments on $\Sone$,
even though the curvatures agree.  Since the transformation
$\zeta_{in}\rightarrow\zeta_1$ is linear, we will obtain the
$\ell^{th}$ and also lower multipoles of $\Sone$.  Similar
considerations apply for the integral over $\Stwo$ as well.

Illustrating the above for the quadrupole moment, we will clearly
obtain a combination of the dipole terms (which vanish for the
individual black holes) and the monopoles.  Putting all of this
together, a straightforward calculation yields:
\begin{equation}
  \begin{alignedat}{2}
      I^{(in)}_2 &{}={}&& \frac{1}{(1+q^2)^2}I_2^{(1)} + \frac{1}{q^2(1+q^2)^2}I_2^{(2)}\\
      &{}+{}&& \sqrt{\frac{5\pi}{4}}\frac{(1-q^2)}{(1+q^2)^2}\left(q^2+\frac{2}{q^2} \right) - \sqrt{20\pi^3}P_2(\zetaneck)\,.
  \end{alignedat}
\end{equation}
The two terms in the first line are the contributions from the
individual quadrupole moments.  In the second line, the first term
consists of the monopole contributions from the two black holes, while
the second term is the contribution from the cusp.  It is
straightforward, if somewhat tedious, to obtain analogous equations
for the higher mass multipole moments.

The considerations thus far are valid only at $\ttouch$ where the scalar
curvature has a $\delta$-function singularity.  As the inner horizon
$\Sin$ evolves away from the cusp, the $\delta$-function is smeared
out and eventually the contributions of the cusp decay.  This however
requires a numerical study which is presented in the next section.

\subsection{Evolution of the mass multipole moments}
\label{subsec:phenmodel}

The above considerations suggest that $\Rin$ can be modeled as the sum
of a smooth function plus $\alpha G$ which limits to a
$\delta$-function at $\ttouch$; the parameter $\alpha$ is the scaling
parameter while $G$ is a function which limits to a $\delta$-function
with unit normalization.  Our strategy will be to first estimate
$\alpha$ and $G$ near the cusp.  Then, $\Rin-\alpha G$ will be a
smooth function on $\Sin$ which is meant to consist of contributions
from the two individial horizons.  We now describe the details of this
fitting procedure.

The neck $\zetaneck$ is a minimum of the scalar curvature:
\begin{equation}
  \left. \frac{d\Rin}{d\zeta}\right|_{\zeta=\zetaneck} = 0 \,\quad \left. \frac{d^2\Rin}{d\zeta^2}\right|_{\zeta=\zetaneck} > 0\,.
\end{equation}
There could be multiple minima of $\Rin$ so this requires an
inspection of $\Sin$ empirically.  If we restrict our attention
sufficiently close to $\ttouch$, this is unambiguously the global
minimum of $\Rin$ over $\Sin$.


\begin{figure*}
  \includegraphics[width=\linewidth]{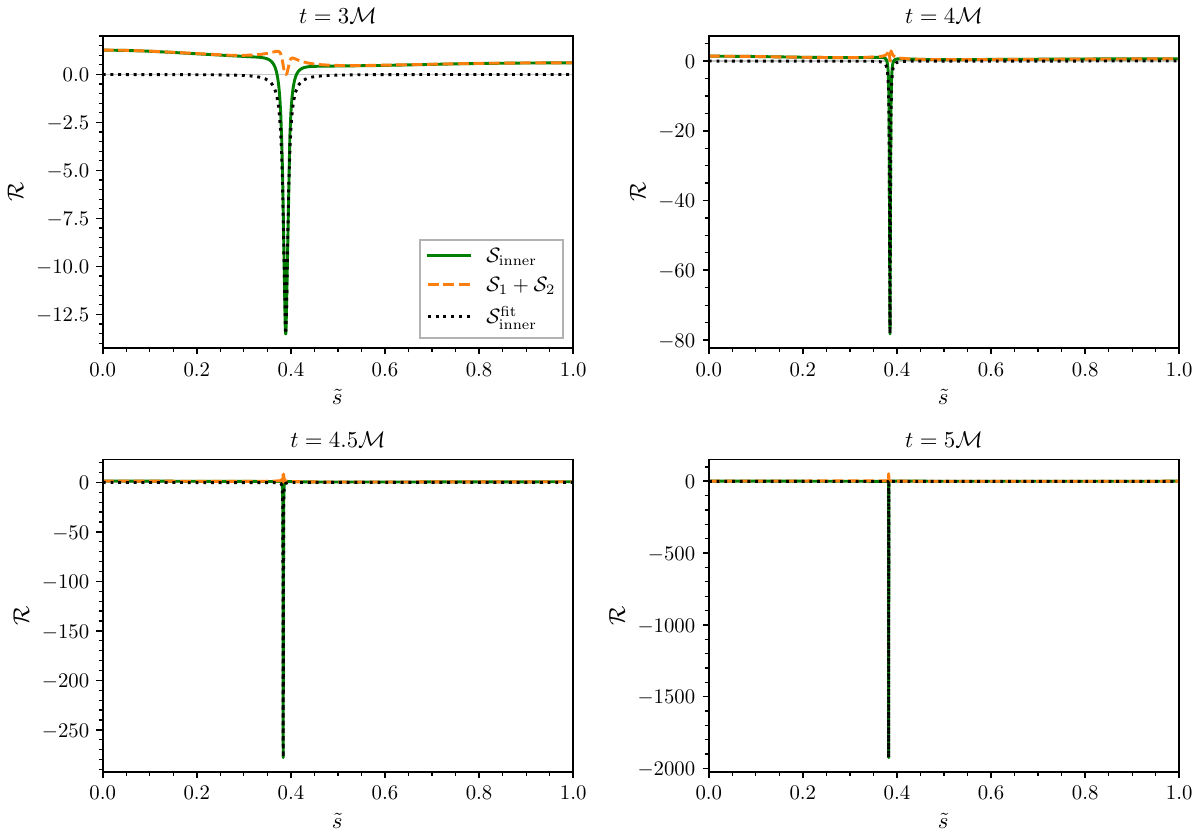}
  \caption{\label{fig:ricci_inner_1_2_1}%
    Ricci scalar $\Rin$ of $\Sin$ (green), the two individual MOTSs
    (orange dashed) and the numerically fitted function (including a
    wrapped Cauchy distribution) to $\Sin$ (black dotted). As
    $\tname\to\ttouch$ the width of the Cauchy decreases and
    approaches a $\delta$-function. The model is meant to capture the
    behavior of $\Rin$ around the neck, and for times near
    $\ttouch$. The behavior near $\zetaneck$ is shown in
    Fig.~\ref{fig:ricci_inner_1_2_2} in better resolve the spike. }
\end{figure*}

\begin{figure*}
  \includegraphics[width=\linewidth]{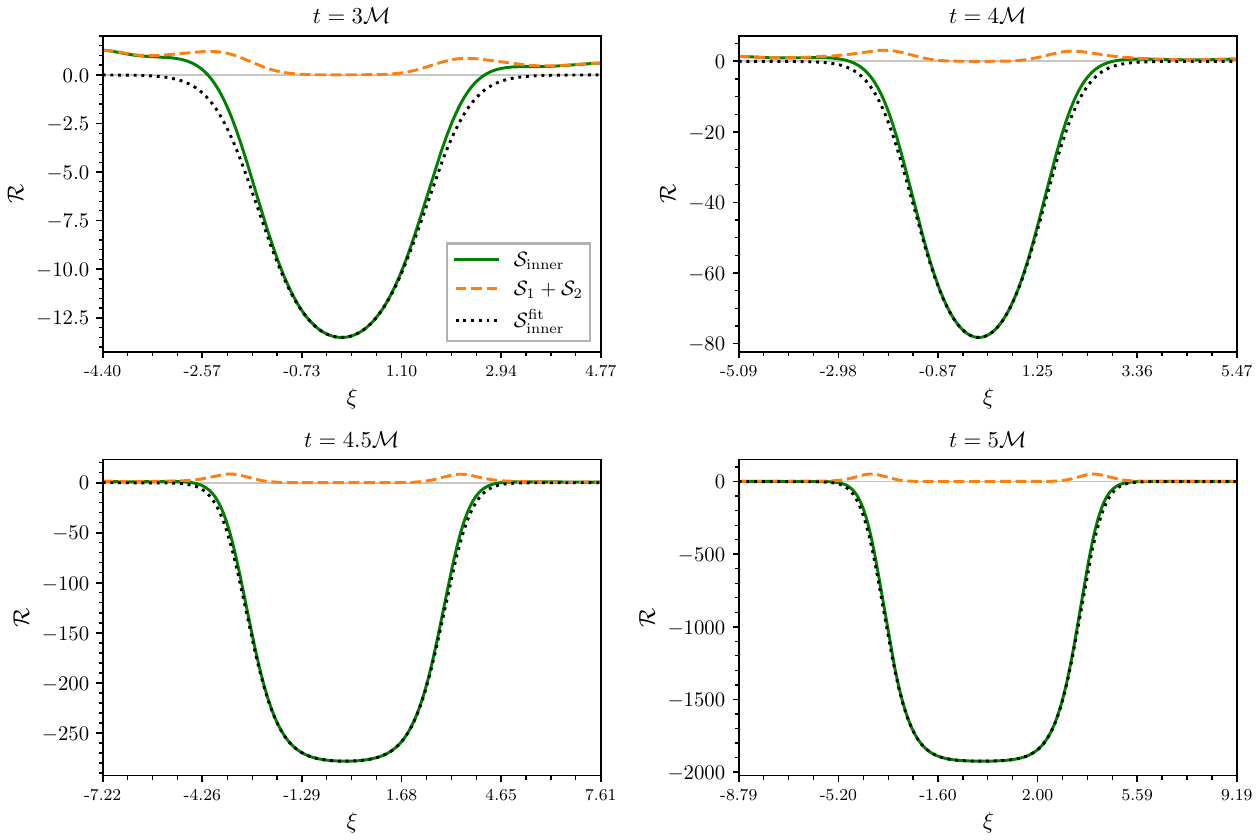}
  \caption{\label{fig:ricci_inner_1_2_2}%
    Ricci scalar $\Rin$ of $\Sin$ (green), the two individual
    MOTSs (orange dashed) and the numerically fitted function (with a
    wrapped Cauchy distribution ) to $\Sin$ (black dotted). As
    $\tname\to\ttouch$ the width of the Cauchy decreases and
    approaches a delta function. In this version of the figure the
    horizontal axis is the transformed coordinate
    $\xi =
      \operatorname{asinh}\!\left(\frac{\stilde-\stilde_{\min}(t)}{w(t)}\right)$,
    where $\stilde_{\min}(t)$ is the location of the minimum of
    $\mathcal{R}$ on $\Sin$ for the given time slice and $w(t)$ is the
    width parameter controlling the amount of horizontal zoom. Thus,
    $\xi$ is a recentered and stretched version of $\stilde$,
    introduced to resolve the increasingly narrow region near the cusp
    as $t\to \ttouch$.}
\end{figure*}


We use the wrapped Cauchy distribution to model the singular part of
$\Rin$:
\begin{equation}
  G(\theta; \theta_{\star},\sigma) :=  \frac{1}{\pi} \frac{\sinh\sigma}{\cosh \sigma - \cos (2\theta -2\theta_{\star})}
\end{equation}
here scaled to the appropriate range of $\theta$. The distribution is normalized to unity
\begin{equation}
  \int_{0}^{\pi} G(\theta; \theta_{\star},\sigma) d\theta = 1
\end{equation}
Furthermore, the peak of the distribution is located at
$\theta_{\star}$ and $\sigma$ is related to the width around
$\theta_{\star}$. We obtain a $\delta$-function when
$\sigma \rightarrow 0$
\begin{equation}
  \lim_{\sigma\rightarrow 0} G(\theta;\theta_{\star},\sigma) = \delta(\theta-\theta_{\star})\,.
\end{equation}

We fit a wrapped Cauchy profile, multiplied by a scaling factor, to the neighborhood of the peak of the data, i.e.\ to the region near the neck. Since the wrapped Cauchy model is intended only as a local description, we do not fit it over the full angular domain. For each time slice, we first locate the peak and construct a small symmetric fitting window around it. We then enlarge this window one point at a time on each side and solve the resulting nonlinear least-squares problem with a Levenberg--Marquardt-type procedure. To improve conditioning in the sharpening regime, we fit in the variables $(q, \alpha)$, where $q=\log \sigma$, rather than in $(\sigma, \alpha)$ directly, thereby enforcing $\sigma>0$ automatically. We also use an algebraically equivalent but numerically more stable form of the wrapped Cauchy kernel, which avoids loss of significance when $\sigma$ becomes very small.

The accepted fitting window is chosen adaptively. We accept the first converged local fit as a baseline and then compare each larger candidate window to the previously accepted one. A larger window is retained only if the fit remains sufficiently stable according to three diagnostics: the change in the fitted width parameter, the relative mean-square residual, and the predictive error at the newly added edge points. In this way, the fit is allowed to grow while it continues to represent the local peak accurately, but it is prevented from absorbing regions where the data are no longer well described by a wrapped Cauchy profile. Near $t_{\mathrm{touch}}$, where the problem becomes very stiff, we allow a small number of rejected enlargements before terminating the growth of the window, so that a single marginal step does not end the procedure prematurely.

\begin{figure*}
  \includegraphics[width=.45\linewidth]{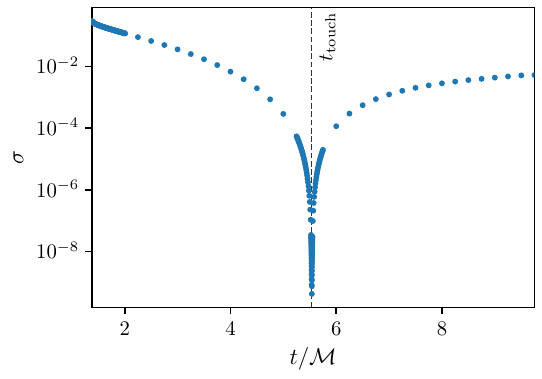}\hfill
  \caption{Fitted values for $\sigma$ parameter of the wrapped Cauchy distribution. The dashed line denotes the time instance where $t=\ttouch$.}
  \label{fig:sigma}
\end{figure*}

\begin{figure*}
  \includegraphics[width=.45\linewidth]{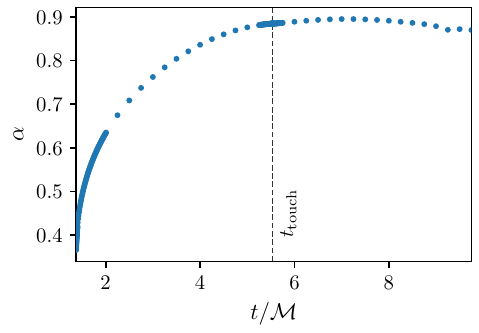}
  \caption{Fitted values for the scaling parameter $\alpha$ of the
    wrapped Cauchy distribution. The dashed line denotes the time
    instance where $t=\ttouch$.}
  \label{fig:scaling}
\end{figure*}

Because of the increasingly singular behavior as $t \to t_{\mathrm{touch}}$, arbitrary-precision arithmetic is used throughout the fitting stage. The fitted values of $\sigma$ decrease rapidly and are consistent with $\sigma \to 0$ as $t \to t_{\mathrm{touch}}$, as can be seen in Fig. \ref{fig:sigma}. The associated scaling parameter is tracked separately as a function of time, as shown in Fig. \ref{fig:scaling}.

\begin{figure*}
  \includegraphics[width=.45\linewidth]{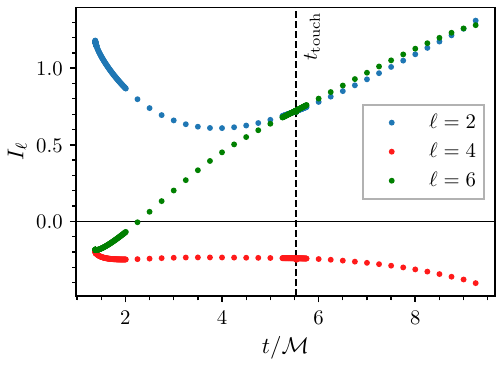}\hfill
  \caption{Multiple moments of $\RR_{12}$ after subtraction of our
    model. The dashed line denotes the time instance where
    $t=\ttouch$.}
  \label{fig:multiple_moments_subtracted}
\end{figure*}

Once the local wrapped Cauchy model has been fitted, we subtract the
fitted inner profile from the full data to define the remainder, which
we interpret as $\mathcal{R}_1+\mathcal{R}_2$. We then compute the
multipole moments $I_\ell$ of this residual field. For the required
angular integrals, we interpolate the sampled data using a Piecewise
Cubic Hermite Interpolating Polynomial (PCHIP), as implemented in
\texttt{DataInterpolations.jl} \cite{Bhagavan2024}, and integrate the
interpolant over the sampled angular interval. This choice preserves
monotonicity near sharp features and avoids the overshoot often
produced by higher-order spline interpolants when the data are not
fully smooth.  The resulting fits for different times are shown in
Fig.~\ref{fig:ricci_inner_1_2_1}, and in greater detail around the
neck in Fig.~\ref{fig:ricci_inner_1_2_2}.  These figure show both the
singular and smooth parts of the fit.  Fig.~\ref{fig:sigma} shows the
width of the Cauchy distribution, which vanishes in the limit
$t\rightarrow\ttouch$ as expected, and Fig.~\ref{fig:scaling} shows
the scaling parameter $\alpha$.  Finally, with the singular part
$\alpha G$ determined by the fit, we can subtract it and obtain the
smooth part corresponding to $\RR_{12}$ and thus it's multipole
moments. This is shown in
Fig.~\ref{fig:multiple_moments_subtracted}. These moments would
eventually asymptote to the multipole moments of the remnant black
hole, and in this case since the final remnant is a Schwarzschild
black hole, it means these will eventually decay away.  This is
studied in greater detail in \cite{Gupta:2018znn,Mourier:2020mwa}.
The moments we have calculated on $\Sin$ need to be matched with those
on the apparent horizon to obtain a complete description.  This will
take us away from the main topic of this work, which is the merger
itself, and will be presented elsewhere.

\section{Conclusions}
\label{sec:conclusions}

A binary black hole merger can be understood as a merger of MOTSs.
During the merger, the two MOTSs $\Sone$ and $\Stwo$ associated with
the two progenitor black holes approach each other and eventually
merge with the inner MOTS $\Sin$ associated with the remnant black
hole.  At the merger time, $\Sin$ is seen to have a cusp. This can be
used not only to connect properties such as mass and multipole moments
of the progenitors with those of the remnant black hole, but to obtain
a detailed record of the time evolution of these parameters.

We have investigated this for head-on collisions of two non-spinning
black holes.  Generally, the time evolution of the mass and mass
multipole moments is smooth except at the time $\ttouch$ where the two
individual MOTSs merge to yield the remnant.  There is a discontinuous
jump at $\ttouch$ corresponding to a $\delta$-function singularity of
the curvature at the cusp.  The magnitude of these discontinuities is
closely related to basic topology of surfaces.  The jump in the mass
results just from the additivity of the area, while the jump in the
higher multipole moments is determined by the Gauss invariant (which
controls the strength of the $\delta$-function singularity in the
curvature).  Away from $\ttouch$, this $\delta$-function is rapidly
``smeared'' over $\Sin$, and is not detectable at late times.  In this
sense, accurate numerical calculations of the MOTSs at $\ttouch$ are
essential for understanding this phenomenon.  While these jumps are
only part of the total change in these quantities starting from the
progenitors at early times to the final remnant black hole, these are
often the bigger effects.  Consider for example the area of the
black holes.  The area, and therefore the irreducible mass generally
increases gradually during the smooth evolution of the MOTSs. On the
other hand the final mass is less than the sum of the initial
masses. This discrepancy is explained by the larger discontinuous
decrease at $\ttouch$.

Looking ahead, the most interesting aspect is to investigate the
effects of angular momentum (and its higher multipoles), and to go
away from axisymmetric configurations.  To illustrate some of these
issues, let us briefly consider angular momentum. For spinning black
holes, the angular momentum is determined by a 1-form
$\widetilde{\omega}_a$ on the MOTS. Integrals of
$\widetilde{\omega}_a$ contracted with suitable axial symmetry vector
fields $\phi^a$ yields the components of the angular momentum vector
$\mathbf{J}$.  At $\ttouch$, we would therefore have the vectors
$\mathbf{J}_1,\mathbf{J}_2$ for the individual horizons $\Sone,\Stwo$
respectively, and $\mathbf{J}$ for $\Sin$.  How would
$\mathbf{J}_{1,2}$ be related with $\mathbf{J}$ at $\ttouch$?  Are
there any topological restrictions on $\widetilde{\omega}_a$ such as
the Gauss invariant?  An important topological property of closed
surfaces is that the space of harmonic 1-forms is connected with the
genus.  Thus on spheres, there exist no harmonic 1-forms.  Does this
fact lead to restrictions on $\widetilde{\omega}$? Moreover, the
notion of angular momentum in classical mechanics depends on the
center of mass; there are additional terms in the angular momentum
when the reference point differs from the center of mass.  We have
seen that the mass-dipole moment at $\ttouch$ is non-vanishing, and
this might be interpreted as a jump in the center of mass. Does this
have any implications for angular momentum?  How would the direction
of $\mathbf{J}$ be related with the directions of $\mathbf{J}_{1,2}$
when they are not aligned?

Next, in the case we have studied in this paper, the presence of
symmetries leads to several simplifications. For example, $\Sone$ and
$\Stwo$ remain exactly axisymmetric and they touch at the poles.  This
will not hold for more general configurations. Finally, an important
physical effect which has been extensively studied is the recoil
velocity produced by the merger.  This has important astrophysical
implications, and very large recoil velocities have been found for
certain configurations.  Would it be possible to obtain a
recoil velocity from these considerations?

Finally we mention possible connections with aspects of binary black
hole waveforms.  Analyses of numerical relativity waveforms have shown
intriguing properties at the merger.  Of relevance to us are the
results in \cite{Borhanian:2019kxt} which analyze different modes of
the gravitational wave signal.  In the inspiral regime, post-Newtonian
theory predicts amplitudes of the various modes in terms of the binary
parameters.  Interestingly, it is shown in \cite{Borhanian:2019kxt}
that these amplitudes are often preserved across the merger.  This
fact then can be used to predict the amplitude of ringdown modes in
terms of the progenitor binary system parameters.  It is plausible
that there is a link between this behavior of the waveforms with those
of the source multipole moments that we have studied here.  The
horizon dynamics of $\Sin$ and especially those of the outermost
horizon also contain imprints of the ringdown modes
\cite{Forteza:2021wfq,Mourier:2020mwa,Khera:2023oyf}.  The connection
between these various complementary aspects of the merger regime could
be useful in gravitational wave astronomy and tests of general
relativity, especially regarding black hole spectroscopy
\cite{Berti:2025hly,Berti:2016lat,Dreyer:2003bv}.

\begin{acknowledgments}
  We are deeply thankful to Lars Andersson, Abhay Ashtekar, Ivan
  Booth, Jos\'e~Luis~Jaramillo, and Ricardo Uribe-Vargas for helpful
  suggestions and fruitful discussions.
\end{acknowledgments}

\appendix

\section{Ricci scalar at the poles}
\label{app:ricci}

We compute here the Ricci scalar of an axisymmetric MOTS $\Surf$ at
its two poles. Dealing with the coordinate singularity requires some
work, which we describe here.  Note that we assume $\Surf$ to be an
immersed sphere such that it necessarily possesses two poles on the
$z$ axis (as opposed to MOTSs of toroidal topology which we do not
consider here).  The steps to arrive at the final expression are very
similar to those taken in Ref.~\cite{Booth:2021sow} and we shall
repeat just the cornerstones.

We work in coordinates $(\rho, \varphi, z)$ related to our Cartesian
coordinates via $\rho^2 = x^2 + y^2$ and $\tan\varphi = y/x$ such
that $\partial_\varphi = -y\partial_x + x\partial_y$ is the
rotational Killing field.
The $(\rho,z)$ half-plane for constant $\varphi$ can then simply be
taken as the $(x>0,z)$ half-plane.
Let $\gamma(s)$ be a curve in this $(\rho,z)$ half-plane
such that $\Surf$ is the surface of revolution of $\gamma$ around
the $z$ axis. We take the parameter $s$ to be the proper length of
$\gamma$ measured from one of the two poles.
The tangent $T := \dot\gamma := \frac{\partial\gamma}{\partial s}$
then has unit length, $T^a T_a = 1$.
Let further $R$ be the circumferential radius defined by
\begin{equation*}
  2\pi R = \int_0^{2\pi} \sqrt{h(\partial_\varphi,\partial_\varphi)}\ d\varphi \,,
\end{equation*}
i.e. $R^2 = h_{\varphi\varphi}$.
Taking now $(s,\varphi)$ as coordinates on $\Surf$, the induced
metric on $\Surf$ becomes
\begin{equation}\label{eq:metricOnS}
  q_{AB} = \begin{pmatrix}
    1 & 0 \\0 & R^2
  \end{pmatrix}\,.
\end{equation}
This form can be used to find a simple expression for the intrinsic
scalar curvature $\RR$ of $\Surf$, i.e. with the Ricci tensor
$R_{AB} := R^C_{\ A\,CB}$ we find
\begin{equation}\label{eq:scal}
  \begin{aligned}
    \RR & = q^{ss} R_{ss} + q^{\varphi\varphi} R_{\varphi\varphi}     \\
        & = 2 q^{ss} R^{\varphi}_{\ s\,\varphi s}                     \\
        & = 2 \left( \partial_s^2 \ln R + (\partial_s\ln R)^2 \right) \\
        & = -2 \frac{\ddot R}{R} \,,
  \end{aligned}
\end{equation}
where as before the dot represents differentiation with respect to
$s$.

The (normalized) left-hand normal to $\gamma$ can be written as (see
again \cite{Booth:2021sow})
\begin{align*}
  N^\flat & = \sqrt{\bar h} (-T^z\ d\rho + T^\rho\ dz) \\
  N       & = \frac{1}{\sqrt{\bar h}} \left(
  -(h_{z\rho}T^\rho + h_{zz}T^z) \partial_\rho
  + (h_{\rho\rho}T^\rho + h_{\rho z}T^z) \partial_z
  \right) \,,
\end{align*}
where $(N^\flat)_a := N_a = \bar h_{ab}N^b$,
$\bar h_{ab}$ is the 2-metric on the half-plane and $\bar h$ its
determinant.
We assume $\gamma$ to be parametrized such that $N$ is pointing in
the outside direction and we will call $\gamma(s=0)$ the
  {\em north pole} of $\Surf$.
With $\bar D_a$ denoting the covariant derivative
compatible with $\bar h_{ab}$, we write the acceleration
\begin{equation*}
  \bar D_T T = \kappa N \,.
\end{equation*}
The expansion of the outgoing and ingoing null normals can then be
expressed as, respectively,
\begin{align*}
  \Theta_{(\ell)} & = k_u + (-\kappa + \bar D_N \ln R)     \\
  \Theta_{(n)}    & = k_u - (-\kappa + \bar D_N \ln R) \,.
\end{align*}
Here, $k_u = q^{ij} K_{ij}$ is the trace of the extrinsic curvature
of $\Surf$ with respect to its timelike normal $u$.
For $\Surf$ to have vanishing expansion, we must hence have
\begin{equation}\label{eq:kappaForMOTS}
  \kappa = \kappa^\pm = k_u \pm \bar D_N \ln R \,.
\end{equation}
If $\kappa = \kappa^+$, then $\Surf$ is a MOTS.
Note that we can reverse the direction of $\gamma$, i.e.
$\gamma(s=0)$ is the {\em south pole} of $\Surf$ and the outward
normal is $-N$. Then, $\Surf$ is a MOTS if $\kappa = \kappa^-$.

To compute the value of $\RR$ at the poles, we let $s\to0$ and use
$\kappa=\kappa^+$ or $\kappa=\kappa^-$ for the north and south pole,
respectively.
To this end, first let
$R_{T} := T^a \bar D_a R$,
$R_{TT} := T^a T^b \bar D_a \bar D_b R$, and so on.
Then
\begin{align*}
  \ddot R = \dot R_T & = T^a \partial_a (T^b \bar D_b R)                             \\
                     & = (T^a \bar D_a T^b) \bar D_b R + T^a T^b \bar D_a \bar D_b R \\
                     & = \kappa R_N + R_{TT}
\end{align*}
As shown in Ref.~\cite{Booth:2021sow}, $\kappa$ remains
finite and nonzero in the limit $s\to0$. However, $R_N$, $R_{TT}$
and $R$ vanish. We hence evaluate this limit with L'Hospital's rule
using
\begin{equation*}
  \begin{aligned}
    \dot R_N    & = -\kappa R_T + R_{TN}     \\
    \dot R_{TT} & = 2\kappa R_{TN} + R_{TTT}
  \end{aligned}
\end{equation*}
and noting that $R_{TN} = 0$ for $s=0$.
With these identities and using \cite{Booth:2021sow}
\begin{equation}\label{eq:kappa0}
  \kappa^\pm_0 = \lim_{s\to0} \kappa^\pm
  = \frac{1}{2} \left( \frac{R_{TN}}{R_T} \pm k_u \right) \,,
\end{equation}
we finally have
\begin{equation}\label{eq:scal0}
  \begin{aligned}
    \RR_0 :={} & \lim_{s\to0} \RR = \lim_{s\to0}\left( -2 \frac{\ddot R}{R} \right)                                 \\
    ={}        & -\kappa_0^\pm \left( \frac{R_{TN}}{R_T} \mp k_u \right) - 2 \frac{2\kappa^\pm R_{TN}+R_{TTT}}{R_T} \\
    ={}        & \frac{1}{2} k_u^2 - 2 \frac{R_{TTT}}{R_T} \,.
  \end{aligned}
\end{equation}
Note that $\RR_0$ does not depend on the direction of the normal,
i.e. it takes on the same value whether $\Theta_{(\ell)} = 0$ or
$\Theta_{(n)} = 0$. In other words, if two arbitrary axisymmetric
MOTSs $\Sone$ and $\Stwo$ share a common point
$p \in \Sone\cap\Stwo$ and $p$ lies on the symmetry axis, then the
scalar curvature $\RR$ of $\Sone$ and $\Stwo$ is equal at $p$.

\section{Derivatives of the Ricci scalar near the poles}
\label{app:ricci_derivatives}

\begin{figure*}
  \includegraphics[width=.45\linewidth]{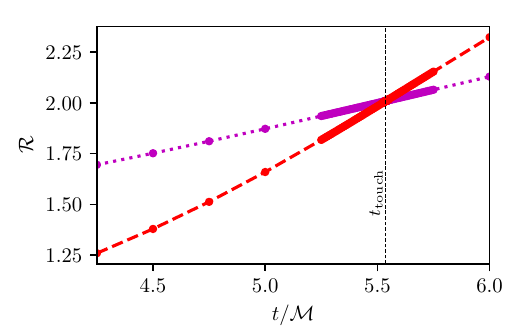}\hfill
  \includegraphics[width=.45\linewidth]{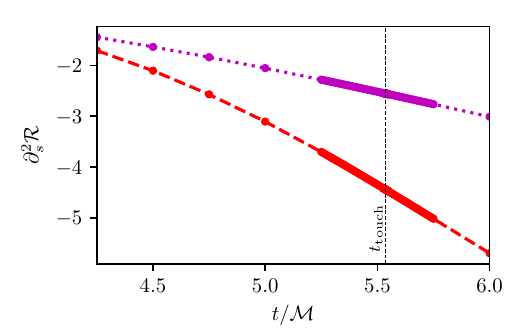}\\
  \includegraphics[width=.45\linewidth]{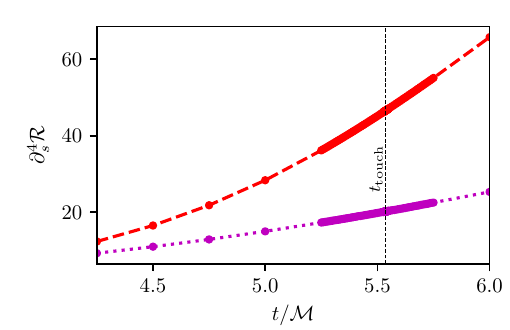}\hfill
  \includegraphics[width=.45\linewidth]{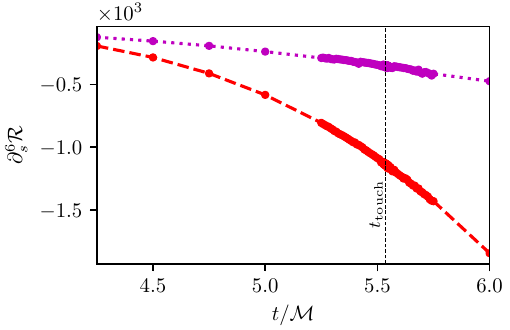}%
  \caption{\label{fig:ricci_compare_theta}%
    Even derivatives of the Ricci scalar $\RR$ of $\Sonetwo$
    evaluated at the facing poles and plotted as a function of time.
    The derivative is taken with respect to the proper length parameter
    $s$.
    The top-left panel shows that the values of $\RR$ coincide at the time
    when $\Sonetwo$ touch. Compare also with
    Fig.~\ref{fig:S12_ricci_delta_plots}.
  }
\end{figure*}

\begin{figure*}
  \includegraphics[width=1\linewidth]{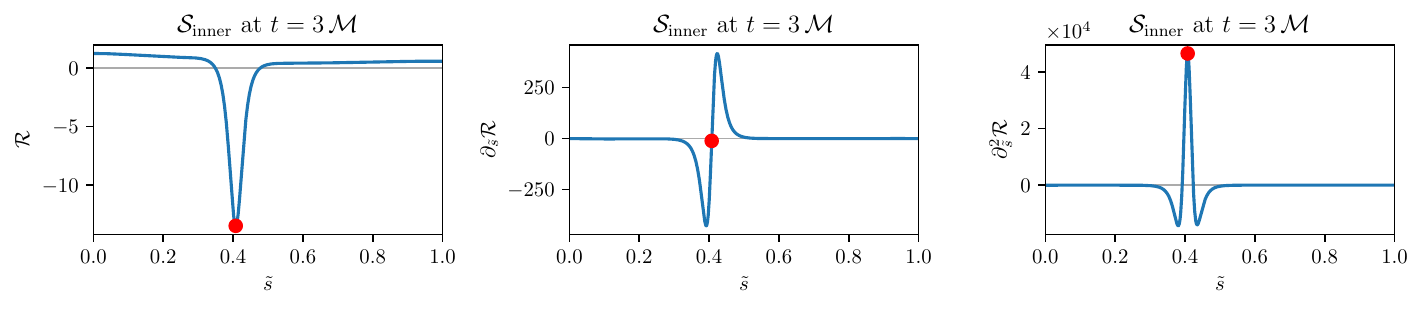}%
  \caption{\label{fig:ricci_inner}%
    Ricci scalar $\mathcal{R}$ of $\Sin$ as function of
    the proper length parameter $\stilde$
    and its first two derivatives.
    The red dot marks the location of the neck.
  }
\end{figure*}

In the main text, we have seen that the mean curvature and the Ricci
scalar of $\Sone$ and $\Stwo$ agree at the point of contact, i.e. at
the north pole of $\Stwo$ and the south pole of $\Sone$.  Furthermore,
the curvatures of $\Sone$ and $\Stwo$ remain smooth at all times, even
at $\ttouch$.  Evidence for this is presented in
Fig.~\ref{fig:ricci_compare_theta}.  In these plots, the Ricci
scalars of $\Sone$ and $\Stwo$ and their first three non-vanishing even
derivatives with respect to proper length $s$ are shown as function of time.
They are evaluated at the facing poles, which coincide exactly at $\ttouch$.
The top-left plot shows again the agreement of the Ricci scalars at
the point of contact.  The remaining three plots are respectively the
second, fourth and sixth derivatives, showing the smoothness of the
Ricci scalars.  Note that the odd derivatives of these functions must
vanish identically due to symmetry.  Finally,
Fig.~\ref{fig:ricci_inner} shows $\Rin$ as function of normalized proper
length $\stilde$ and its first two derivatives.

\bibliography{bibliography}{}

\end{document}